**Effects of forest fire severity on terrestrial carbon emission and ecosystem production in the Himalayan region, India**

**Srikanta Sannigrahi[a], Sandeep Bhatt[b*], Shahid Rahmat[a], Virendra Rana[b], Suman Chakraborti[c]**

[a] Department of Architecture and Regional Planning, Indian Institute of Technology Kharagpur-721302, India

[b] Department of Geology & Geophysics, Indian Institute of Technology Kharagpur-721302, India

[c] Center for the Study of Regional Development (CSRD), Jawaharlal Nehru University, New Delhi 110067, India

*Corresponding author: Sandeep Bhatt

Phone: +917583947841

E-mail: (Srikanta Sannigrahi) : srikanta.arp@iitkgp.ac.in

(Sandeep Bhatt*) : sandeep@gg.iitkgp.ernet.in

(Shahid Rahmat) : shahidrahmat82@gmail.com

(Virendra Rana) : rana.virendra3@gmail.com

(Suman Chakraborti) : suman87_ssf@jnu.ac.in

**Abstract**

Remote sensing techniques have been used effectively for measuring the overall loss of terrestrial ecosystem's productivity and biodiversity due to forest fires. The current research focuses on assessing the impact of forest fire severity on terrestrial ecosystem productivity using different burn indices in Uttarakhand, India. Satellite-based Land Surface Temperature (LST) was calculated for pre-fire (2014) and fire years (2016) using MODerate Resolution Imaging Spectroradiometer (MODIS) to identify the burn area hotspots across all eco-regions in Uttarakhand. In this study, spatial and temporal changes of different vegetation status and burned area indices viz. Normalized Burn Ratio (NBR), Burnt Area Index (BAI), Normalized Multiband Drought Index (NMDI), Soil Adjusted Vegetation Index (SAVI), Global Environmental Monitoring Index (GEMI), Enhanced Vegetation Index (EVI), Normalized Difference Vegetation Index (NDVI) were estimated for both fire and pre-fire years to analyze its relation with ecosystem productivity and associated changes. Additionally, two Light Use Efficiency (LUE) models: Carnegie- Ames-Stanford-Approach (CASA) and Vegetation Photosynthesis Model (VPM) were selected to quantify the terrestrial Net Primary Productivity (NPP) in pre-fire and fire years across all biomes of the study area. The results revealed a statistically significant (at P = 0.05 and P = 0.01) positive correlation between burn indices and estimated change in ΔEVI and ΔNPP (r = 0.54), ΔNDVI and ΔNPP: (r = 0.55), ΔNBR and ΔNPP: (r = 0.36), ΔSAVI and ΔNPP: (r = 0.16), ΔGEMI and ΔNPP (r = 0.16) whereas, a negative trend is reflected between the ΔNMDI and ΔNPP: (r =−0.39), and ΔLST and ΔNPP during the both studied years. In addition, the ΔNPP is highly correlated with the forest fires density (FFD) ($R^2$ = 0.75, RMSE = 5.03 gC $m^{-2}$ $month^{-1}$). The present approach appears to be promising and has a potential in quantifying the loss of ecosystem productivity due to forest fires. A detailed field observation data is required for further training, and testing of remotely sensed fire maps for future research.

## 1. Introduction

Forests are basic components of the global carbon cycle, and forest fires are a serious threat to indigenous forests that degrade net primary productivity (NPP), gross primary productivity (GPP) and carbon sequestration services (Dixon et al., 1994). According to Intergovernmental Panel on Climate Change (IPCC, 1992), burning of forest biomass produce a significant amount of $CO_2$, which is 10% of the annual global methane and 10-20% of the global $N_2O$ emissions leading to the change in atmospheric chemistry in the long run. Forest fires perturb the carbon sequestration capacity of a green canopy and dismantle the surface energy balance of the ecosystem by emitting several greenhouse gases which affecting the spatiotemporal dynamics of carbon pools, thermal properties of regional/global climate and Net Ecosystem Productivity (NEP) (e.g., Flannigan et al., 2000; Amiro et al., 2001; Amiro et al., 2003). Several environmental indicators, including NPP, GPP, NEP etc. has been used widely to analyze the anthropogenic and climatic effects on ecosystem on the local and global scale.

NPP is the net amount of carbon fixed by a green canopy from the atmosphere in a given space and time through photosynthesis and plant respiration (Potter et al., 1993). It is one of the most understood ecosystem processes to analyse the anthropogenic and climatic effects on an ecosystem (Potter et al., 1993; Chu et al., 2016). In addition to this, NPP is useful in capturing vegetation changes across the world and is deployed to track the unprecedented modifications in different biomes (Field et al., 1995; Brouwers and Coops, 2016). Presently, NPP is highly favored as an ecosystem indicator for measuring the capacity of an ecosystem to act as a carbon source or sink (Amiro et al., 2001).

The remotely sensed data has been widely utilized in the past few decades to extract the valuable information about the dynamics of terrestrial ecosystem productivity and vegetation phenological pattern from coarser to finer scales (Field, 1998). The availability (free data) of remote sensing data makes it suitable for estimating extent and monitoring of fire events in the developing countries under scarce data conditions. Over the last few decades, several remote sensing based spectral indices have been explored around the globe to assess the impact of fire on forest ecosystem at local, regional and global scales (Milne, 1986; White et al., 1996; Lentile et al., 2006; Chen et al., 2011; Chen et al., 2015). Different indices have their strengths and shortcomings for assessing and mapping the forest fire severity. Normalized Difference Vegetation Index (NDVI) is the most commonly used index that detects the green vegetative cover (Chuvieco et al., 2004) but it is sensitive to attenuation from atmosphere and aerosols (Carlson and Ripley, 1997). Soil-Adjusted Vegetation Index (SAVI) is modified NDVI with reduced soil background effect making it very sensitive to discriminating vegetation amount in sparsely vegetated areas (Huete 1988; Chenhbouni et al., 1994). The Enhanced Vegetation Index (EVI) has improved sensitivity in high biomass regions and improved vegetation monitoring through a de-coupling of the canopy background signal with less atmospheric influences (Jiang et al., 2008). The Global Environmental Monitoring Index (GEMI) is sensitive in discriminating burned area and is least affected by soil variations, atmospheric variations, and illumination conditions than NDVI (Pinty and Verstraete 1992; Pereira 1999). Besides, the Burned Area

Index (BAI) easily discriminate the burned and low reflectance areas depending on the temporal performance of behavior charcoal after forest fires (Martin, 2006). Normalized Multiband Drought Index (NMDI) uses the difference between two liquid water absorption bands (1.64 mm and 2.13 mm), as the soil and vegetation water sensitive band. Strong differences between two water absorption bands in response to soil and leaf water content has the potential to estimate water content of both soil and vegetation. Therefore, this improved drought index is expected to offer more accurate assessments of drought severity and fire conditions (Wang and Qu, 2007; Wang et al., 2008). Normalized Burn Ratio (NBR) combines the information on the near-infrared (NIR) band centred at approximately 0.8 mm and a shortwave infrared (SWIR) band centred at approximately 2.1 mm to map the burned areas and burn scar (Key and Benson, 1999, 2005; Miller and Yool, 2002; Cocke et al., 2005). As, NIR and SWIR spectral bands have the most change among reflective spectral bands (White et al., 1996; Wagtendonk et al., 2004), therefore, NBR would be one of the most discriminating for burn effects during forest fires. Apart from the mentioned spectral burn indices, the land surface temperature (LST) that uses the thermal bandwidth to detect water, energy interaction between earth and atmosphere, is used as a significant burn indicator to assess the forest fire destruction (Zheng et al., 2016). The available burn indices alone do not provide a sound mapping of forest fires due to atmospheric disturbances and a different number of bands used for their quantification. Therefore, it is recommended to use more than one burn indices to overcome the bias and achieve reliable quantification of forest fires for better assessment of net ecosystem productivity and carbon emission. All the formulaes and spectral bands correspond to the indices used in this study are explained in the Appendix section.

In India, forest fires (natural and anthropogenic) incur an annual loss of ~ \$70 million supporting and regulatory ecosystem service affecting ~ 55% of forest cover (Jha et al., 2016). The state of Uttarakhand (Northern India) witness recurring wildfires (natural and anthropogenic) in its pine forests and grazing lands annually during the dry summer months (Bhandari et al., 2012) that contribute to high carbon emissions and loss of ecosystem services. During 24th April 2016 to 3rd May 2016, the state (Uttarakhand) witnessed severe forest fire events (~ 1600 active forest fires), which destroyed nearly 7.35% (2,166 $km^2$) of the total (24,240 $km^2$) forest cover across 13 different districts of the state (Jha et al., 2016). The aforementioned extreme event took place on 24th of April 2016, and it took nearly 10 days to control (3rd May 2016) the forest fire in the nearby areas. The primary objective of this study is to investigate the impact of forest fire severity on terrestrial ecosystem productivity of a highly sensitive and ecologically important area (Uttarakhand, India). We evaluate the dynamics of terrestrial ecosystem productivity and vegetation phenological pattern using remotely sensed data. This study proposes a novel approach ($\Delta$NPP /$\Delta$burn indices) to quantify the effects of the forest fire severity on terrestrial carbon emission and ecosystem using several indicators (NDVI, SAVI, EVI, NBR, NMDI, GEMI, and BAI) for mapping forest fire in the state of Uttarakhand. The current research also assesses the (1) forest fire severity using selected burn indices for two experimental years (2014: pre-fire and 2016: fire) (2) spatiotemporal behaviour of net primary productivity for pre-fire and fire years (3) impact of forest fire severity on terrestrial ecosystem productivity and carbon emission.

## 2. Materials and methods

### *2.1 Study area*

The Uttarakhand state lies to the south of the Himalayas, comprising of 13 districts that occupying an area of 53,484 $km^2$ (93% hilly area with 65% natural vegetation) (Negi et al., 2009). The topography varies significantly with an elevation ranging between 200 -7800 m above mean sea level (MSL) in the Gangetic plains and the Himalayan region, respectively depicting the high topographical as well as the ecological diversity of the region (**Fig. 1**). The local climate is characterized by sharp variation in temperature and precipitation difference viz. temperature ranges from 40ºC in April-May (Max) and <0ºC in January-February (Min) and annual average precipitation from 1000mm to 1600mm. The vegetation distribution is mostly due to the altitudinal variation (Mishra and Chaudhuri, 2015). Alpine shrubs and meadows, Temperate West-Himalayan broadleaf forest, Himalayan Subtropical Pine forest (susceptible to fire) occupy the areas with an elevation ranging from 3000m to 4800m, 1500m to 2600m and ~1500m above MSL, respectively (Singh and Singh, 1987).

## 2.2 Methods

### *2.2.1 Estimation of Land Surface Temperature (LST)*

In this study, MODIS daily 1km level-3 land surface temperature and emissivity product (MOD11A1) version 5 were extracted from https://lpdaac.usgs.gov/dataset_discovery/modis/modis_products_table/mod11a1_v006 to estimate the daily LST for the pre-fire year (2014) and fire year (2016), respectively (**Table 1**). The two thermal bands, i.e., 31(10.78 – 11.28 μm) and 32 (11.77 – 12 μm) are generally used to calculate daily LST using Split-Window algorithm (Qin et al., 2001; Wan et al., 2004). Here we have used the MOD11A1 daily LST data products for two reference days ($27^{th}$ April and $29^{th}$

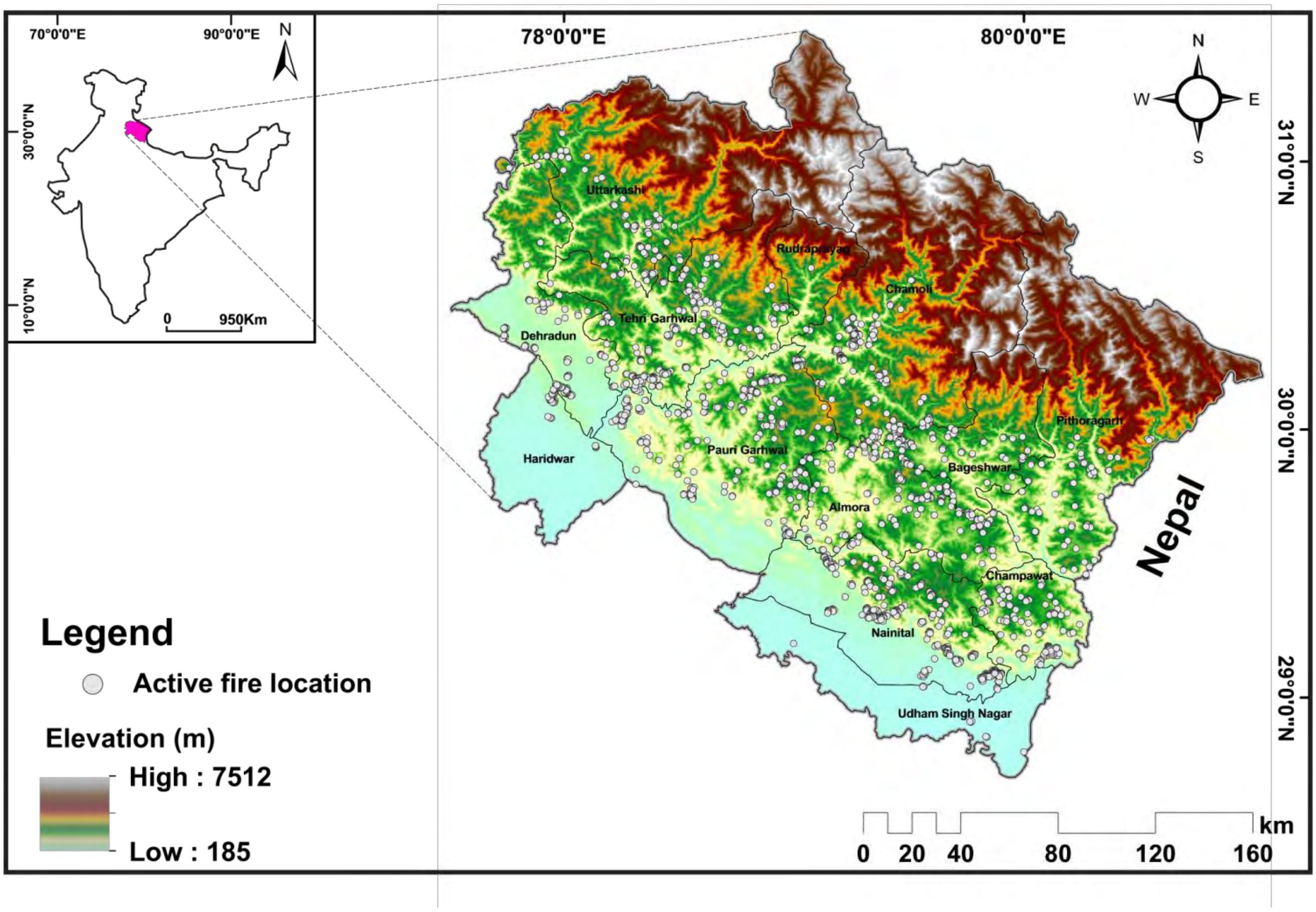

**Fig. 1:** Location of the study area showing the Digital Elevation Model (DEM) of Uttarakhand with active fire locations (source: http://bhuvan.nrsc.gov.in/bhuvan_links.php, MODIS TERRA & VIIRS satellite products).

**Table. 1** Description of the datasets used in the study

| Dataset | Month of Acquisition | Description | Source | Spatial scale | Temporal Scale |
|---|---|---|---|---|---|
| ***MODIS*** | | | | | |
| MOD11A1 | April/May | LST, Emissivity | NASA | 1km | Daily |
| MOD09A1 | April/May | Surface reflectance | NASA | 500m | 8 day |
| MOD13Q1 | April/May | EVI, NDVI | NASA | 250m | 16 day |
| MOD17A2 | April/May | GPP | NASA | 500m | 8 day |
| MOD15A2 | April/May | fPAR/ LAI | NASA | 1km | 8 day |
| MOD09Q1 | April/May | Surface reflectance | NASA | 250 m | 8 day |
| ***LULC database*** | | | | | |
| MCD12Q2 | | Global LULC | NASA | 500m | yearly |
| GLC 2000 | | Global LULC | ESA | 300m | ----------- |
| ***Daily Meteorology*** | | | | | |
| POWER LARC | April/May | Global Agro-climatology data | NASA | $1^{o} \times 1^{o}$ | daily |
| CFSR | April/May | SWAT data | TAMU | $0.3^{o} \times 0.3^{o}$ | hourly |

April) depending on the quality of the data to compute the delta LST between fire and pre-fire years.

### *2.2.2 Estimation of vegetation dynamics*

Two main satellite-based vegetation indices namely, NDVI and EVI, were selected for assessing the vegetation dynamics and to quantify their impact corresponding to the terrestrial ecosystem productivity in pre-fire (2014) and fire years (2016). The 16 days composite product from MODIS (MOD13Q1) version 6 (https://lpdaac.usgs.gov/dataset_discovery/modis/modis_products_table/mod13q1_v006) was used to estimate time series EVI and NDVI for both experimental years. The quality of the targeted pixel's was assured by the given quality flag information of the aforementioned products; however, the bad quality pixels were discarded from the input data used in the study. Furthermore, the aforementioned vegetation indices were retrieved using the ratio of different spectral bands (see appendix).

### *2.2.3 Estimation of burn indices*

The level 3 (500m gridded; 8-days) with 1-7 spectral band surface reflectance products (MOD09A1)(https://lpdaac.usgs.gov/dataset_discovery/modis/modis_products_table/mod09a1_v006) version 6 were used to estimate the burn indices for the forest fire disturbance analysis (Arnett et al., 2015). A defined scale factor equal to 0.0001 for bands 1-7 was used to retrieve the actual pixel information for further analysis. The selected burn indicators were extracted by using the expressions provided in the appendix.

### *2.2.4 Quantification of Net Primary Productivity (NPP) and Carbon emission*

In order to have reliable estimates of NPP (gC $m^{-2}$ $month^{-1}$), two different ecosystem models were used in the current research. Basic information corresponding to the selected models for NPP estimation is included in the manuscript, whereas the mathematical concept is provided in the appendix.

#### *2.2.4.1 NPP estimation using Vegetation Photosynthesis Model (VPM)*

The VPM model, developed by Xiao et al. (2004), is based on the conceptual partitioning of non-photosynthetic vegetation (NPV; mostly senescent foliage, branches, and stems) and photosynthetic vegetation (PAV; mostly chloroplast) within the leaf and canopy. This model is driven by temperature stress scalar, moisture stress scalar and the age of phenology, respectively (see appendix). The estimated Gross Primary Productivity (GPP; gC $m^{-2}$ $month^{-1}$) was converted to the NPP (GPP * 0.53) and then NPP to biomass (gC $m^{-2}$ $month^{-1}$) (NPP * 2.22) (Zhang et al., 2009; Li et al., 2014) for the subsequent analysis.

#### *2.2.4.2 NPP estimation using Carnegie-Ames-Stanford Approach (CASA) Model*

The Carnegie-Ames-Stanford Approach (CASA) model (Potter et al., 1993) is used to estimate the terrestrial NPP by utilizing the satellite imagery information and climatic measurement across the various eco-regions. The net photosynthetic radiation (PAR; MJ $m^{-2}$

year$^{-1}$), the biophysical dynamics (NDVI), and different climatic and environmental stress regulators control the NPP of any biome ($T_{s_1}, T_{s_2}, W_S$) (Potter et al., 1993) (see appendix).

MOD17A2 8 days GPP (https://lpdaac.usgs.gov/dataset_discovery/modis/modis_products_table/mod17a2h_v006) version 6 was used in the current study where simply the rates of change of NPP corresponding to a particular index (Eq. 1-6). They were used to identify the sensitivity between NPP and the selected burn indices thereby making them explicable to the readers. All the required input variables were rescaled into 500m spatial resolution using bilinear interpolation method for the subsequent analysis. Normality of data points was assessed through Shapiro–Wilk test and Kolmogorov–Smirnov test. It was found that data are non-normally distributed. This result was very obvious because of the intensity of fire points was found very low to extremely high across the study region. Pearson correlation coefficient and Spearman rank correlation (non-parametric) analysis was performed to analyze the relationship between NPP and the selected burn indices. Total 2048 sample points (active fire locations, shown in Fig. 1) were utilized for the analysis. **Table. 1** provides the description of the data type, data source, spatial and temporal extent of the data set used in the current research.

$$\Delta NPP = NPP_{prefire} - NPP_{fire} \qquad (1)$$

$$\Delta(NPP / \mathrm{BAI}) = \frac{\Delta NPP}{\Delta BAI} \qquad (2)$$

$$\Delta(NPP / NBR) = \frac{\Delta NPP}{\Delta NBR} \qquad (3)$$

$$\Delta(NPP/\mathrm{SAVI}) = \frac{\Delta NPP}{\Delta SAVI} \quad (4)$$

$$\Delta(NPP/\mathrm{NMDI}) = \frac{\Delta NPP}{\Delta NMDI} \quad (5)$$

$$\Delta(NPP/\mathrm{GEMI}) = \frac{\Delta NPP}{\Delta GEMI} \quad (6)$$

Where $\Delta NPP$ is the $NPP$ (gC m$^{-2}$ month$^{-1}$) difference between the pre-fire and fire years, respectively. In addition, ecosystem light use efficiency (ELUE; gC MJ$^{-1}$) was quantified directly from the estimated GPP (Ma et al., 2014).

$$ELUE = \frac{GPP}{PAR} \quad (7)$$

Where $PAR$ is the photosynthetically active radiation (MJ m$^{-2}$ month$^{-1}$). Subsequently, the spatial coherence of the two ecosystem models (CASA & VPM) and the sensitivity between burn indices and NPP were evaluated by the standard model validation technique (Ma et al., 2014).

$$R^2 = 1 - \frac{\sum (\mathrm{x}_i - y_i)^2}{\sum y_i^2 - \frac{y_i^2}{N}} \quad (8)$$

Where $R^2$ is the coefficient of determination, $\mathrm{x}$ and $y$ are the explanatory and response variables of the $i^{\mathrm{th}}$ month, $N$ is total number of samples.

### *2.3 Quantification of greenhouse gas emissions*

The spatio-temporal emission of the major greenhouse components (C, $CO_2$, $CH_4$, $N_2O$, $NO_x$ and Particulate matter) was estimated through NPP for the pre-fire and fire years. NPP assess the environmental impact of forest fire and associated loss of natural resources in a highly enriched ecosystem.

***a) Release of carbon***

Intergovernmental Panel on Climate Change (IPCC) guidelines for national greenhouse inventories (IPCC, 1996; 2006) were followed in calculating greenhouse gas emissions due to forest fires as given below:

$$C = \Delta Biomass \times 0.9 \times 0.45 \qquad (9)$$

Where $C$ (g C) is the amount of carbon released due to forest fire; $\Delta Biomass$ is the changes in biomass between the pre-fire and fire years; $0.9$ represents the fraction of biomass oxidized on site and $0.45$ represent the actual carbon content (IPCC, 2006; Yan et al., 2009; Meinshausen et al., 2009).

The amount of gaseous carbon (g $CO_2$, $CH_4$, CO) compounds emission retrieved as follows:

$$E_j = \varepsilon_j \times \delta_j \times C \qquad (10)$$

Where $\varepsilon_j$ is the fraction of total carbon emitted as compound j and $\delta_j$ is the fraction of passage from the emission of carbons to the emission of the specific compound. The $\varepsilon_j$ and $\delta_j$ values for $CO_2$, $CH_4$, CO is considered as 0.888, 0.012, 0.1 and 3.67, 1.33, 2.33, respectively (IPCC, 1996; 2006; Yan et al., 2009; Meinshausen et al., 2009).

***b) Release of Nitrogen Compounds***

Emissions of Nitrogen compound (g $NO_2$, $NO_x$) were quantified as follows:

$$N = \gamma^{*} C \quad (11)$$

$$E_j = \varepsilon_j \times \delta_j \times N \quad (12)$$

Where $\gamma^{*}$ is the proportion of emitted Carbon and Nitrogen (0.01), the values for the coefficients $\varepsilon_j$ and $\delta_j$ for $NO_2$ and $NO_x$ are specified as 0.007, 0.012 and 1.57, 2.14, respectively (IPCC, 1996; 2006; Yan et al., 2009; Meinshausen et al., 2009).

## 3. Results and Discussion

### *3.1 Spatio-temporal dynamics of burn indices and its usability on forest fire disturbance*

The forest fires in Uttarakhand occurred during the prolonged dry summer, when mean atmospheric temperature exceeded the normal level (Mukhopadhyay, 2001). A detailed land use/land cover (LULC) map with actual forest fire locations and estimated forest fire intensities of different districts in Uttarakhand (**Fig. 2a, b**) reveals maximum spots lie in moist deciduous and subtropical pine forest ecosystem (Jha et al., 2016). The occurrence of high fire intensity at low altitude (≤ 1500 m above MSL) can be attributed to plant species (e.g., *Pinus roxburghii, Quercus leucotrichophora)*, and proximity to the villages that make them susceptible to anthropogenic interferences (e.g., clearance of forest cover, stimulating grazing intensity, dispersing plant communities and dismantling plant functional traits, changing ignition patterns, etc.) (Bowman et al., 2011; Balch et al., 2017; Chand et al., 2007; Sharma et al., 2017; Kumar and Ram, 2005). Moreover, the high inflammability of igniting material of Pine forest is depend on low moisture content, and the high ambient temperature is increasing the dryness of fuel loads lying on the forested strand promoting high-density forest fires in summer time (Sharma et al., 2011). The abundance of dry leaves in forest strand and windward face of the surface topography could be a plausible reason for gaining a relatively higher proportion of available surface energy to trigger the acute forest fires in 13 districts of Uttarakhand. The selected burn indices (NBR, BAI, EVI, LST, GEMI, NDVI, SAVI, and NMDI) have been employed to quantify the spatially explicit changes of ecosystem production caused by the 2016 forest fire in Uttarakhand.

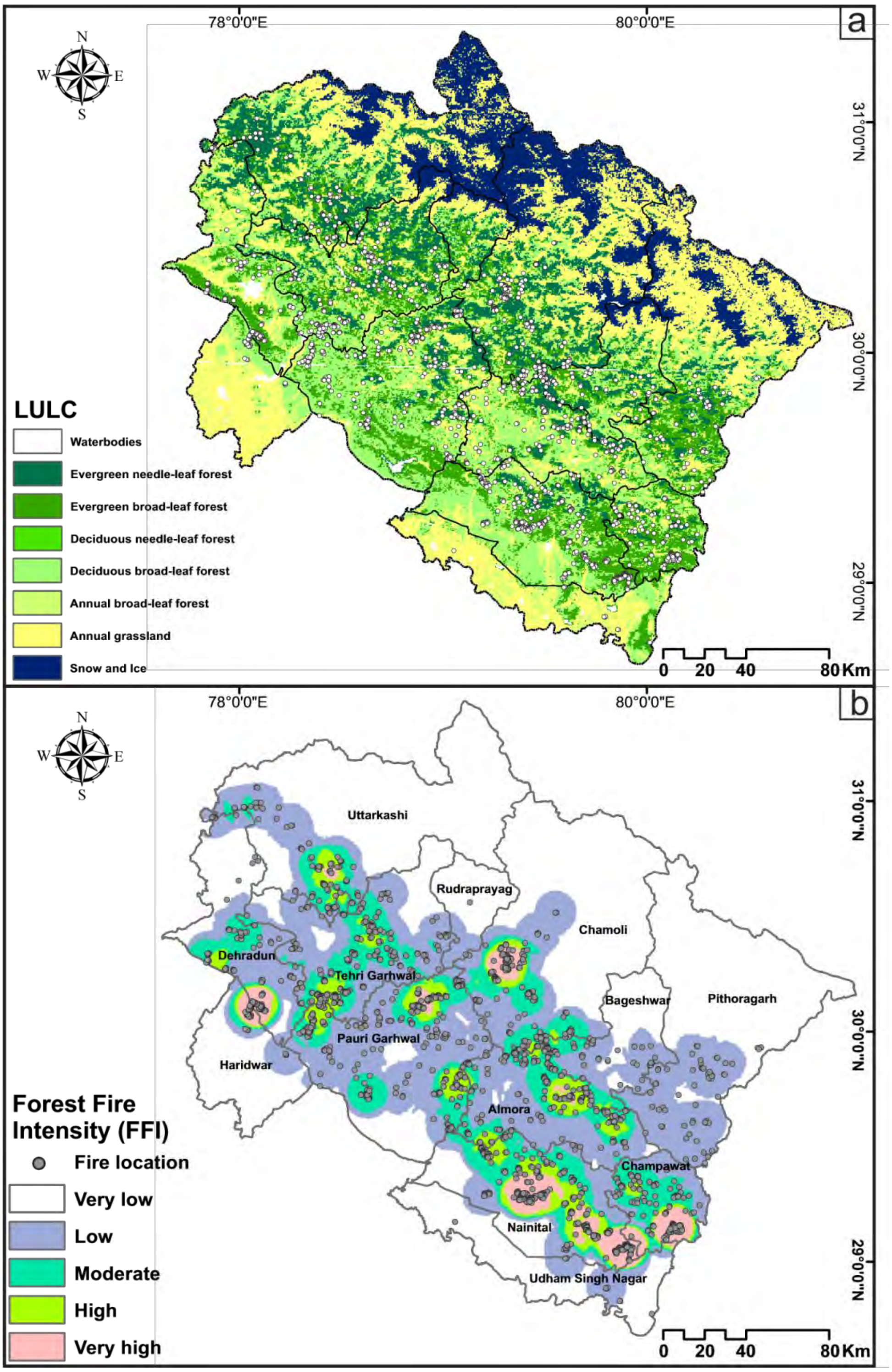

**Fig. 2:** (a) Detailed Land-Use and Land-Cover (LULC) distribution with active forest fire locations; (b) The distribution of the Forest Fire Intensity (FFI) in the region.

The district level satellite-derived LST (for the same dates) in the pre-fire (2014) and fire years (2016) reveals the spatiotemporal discrepancies and unusual changes in the surface temperatures of the different districts caused by the severe forest fires. The maximum LST in the pre-fire year (27th April 2014) was 44.97ºC which augmented to 51.83ºC in the fire year (27th April 2016), whereas the change in temperature was more prominent on 29th April (pre-fire and fire years), i.e., more than 8ºC. For April 27, the maximum change in LST (ºC) between the pre-fire and fire years was observed in Haridwar (5.34), followed by Pithoragarh (4.06), Udham Singh Nagar (3.98), Nainital (3.46), Dehradun (3.07), Chamoli (2.75), respectively. While the districts of Rudraprayag (-0.74), Bageshwar (-0.63), and Almora (-0.09) exhibit minimum to negligible change in LST. For April 29, the delta LST (ºC) values is found maximum for Uttarkashi (9.8), followed by Dehradun (6.77), Udham Singh Nagar (6.56), Tehri Garhwal (5.59), Haridwar (5.48), Rudraprayag (4.57), and Nainital (4.55). (**Table. S1; see supplementary file**).

**Table. S1** Districts wise variation of LST (ºC) between the pre-fire and forest fire years in the study region.

| Districts | 27th April 2014 (Pre-fire) | | 27th April 2016 (Fire) | | ∆LST (ºC) | | 29th April 2014 (Pre-fire) | | 29th April 2016 (Fire) | | ∆LST (ºC) | |
|---|---|---|---|---|---|---|---|---|---|---|---|---|
| | **Mean** | **Max** | **Mean** | **Max** | **Mean** | **Max** | **Mean** | **Max** | **Mean** | **Max** | **Mean** | **Max** |
| Rudraprayag | 22.73 | 36.61 | 21.99 | 37.99 | -0.74 | 1.38 | 19.84 | 33.49 | 24.41 | 36.87 | 4.57 | 3.38 |
| Tehri Garhwal | 29.89 | 39.71 | 31.97 | 45.65 | 2.08 | 5.94 | 27.31 | 38.17 | 32.9 | 43.77 | 5.59 | 5.6 |
| Udham Singh Nagar | 38.43 | 44.97 | 42.41 | 46.69 | 3.98 | 1.72 | 34.92 | 39.31 | 41.48 | 49.41 | 6.56 | 10.1 |
| Almora | 35.51 | 41.73 | 35.42 | 43.55 | -0.09 | 1.82 | 33.89 | 40.69 | 37.11 | 44.69 | 3.22 | 4 |
| Bageshwar | 26.98 | 35.89 | 26.35 | 35.81 | -0.63 | -0.08 | 24.11 | 31.91 | 27.91 | 35.79 | 3.8 | 3.88 |
| Chamoli | 17.45 | 37.85 | 20.2 | 38.39 | 2.75 | 0.54 | 16.94 | 34.49 | 20.09 | 39.73 | 3.15 | 5.24 |
| Champawat | 34.45 | 39.25 | 34.54 | 42.85 | 0.09 | 3.6 | 32.28 | 35.63 | 35.77 | 43.19 | 3.49 | 7.56 |
| Dehradun | 34.73 | 40.85 | 37.8 | 48.73 | 3.07 | 7.88 | 32.53 | 41.89 | 39.3 | 48.57 | 6.77 | 6.68 |
| Haridwar | 37.37 | 41.65 | 42.71 | 51.83 | 5.34 | 10.18 | 36.41 | 41.25 | 41.89 | 49.51 | 5.48 | 8.26 |
| Nainital | 35.9 | 43.01 | 39.36 | 50.27 | 3.46 | 7.26 | 35.06 | 39.85 | 39.61 | 50.01 | 4.55 | 10.16 |
| Pauri Garhwal | 35.44 | 43.65 | 37.25 | 49.11 | 1.81 | 5.46 | 34.62 | 40.27 | 37.93 | 50.11 | 3.31 | 9.84 |
| Pithoragarh | 19.19 | 38.73 | 23.25 | 40.33 | 4.06 | 1.6 | 18.44 | 35.57 | 22.42 | 38.75 | 3.98 | 3.18 |
| Uttarkashi | 16.68 | 36.03 | 18.35 | 42.21 | 1.67 | 6.18 | 14.34 | 31.91 | 24.14 | 45.01 | 9.8 | 13.1 |
| **Average** | **29.60** | **39.99** | **31.66** | **44.11** | **2.07** | **4.11** | **27.75** | **37.26** | **32.69** | **44.26** | **4.94** | **7.00** |

There are several factors responsible for rising of surface and air temperature in the study region. The weather record of 2015 reveals that the maximum rainfall in August with a yearly average of 1138mm was 47% less than the average precipitation received in the last century. This had enhanced the moisture deficit conditions of the state in the early summer of 2016 (Jha et al., 2016; Sharma and Pant, 2017). The weak Westerlies (contributed around 20% of total rainfall in Uttarakhand) and associated below normal winter precipitation is also responsible for the catastrophe fire event in 2016 in Uttarakhand (Sati and Juyal, 2016). The late monsoon precipitation is utmost crucial for sustaining soil moisture and prevention of fire intensities especially in summer times (Jha et al., 2016; Sati and Juyal, 2016). Additionally, this region has experienced an unusual winter (February) forest fire incidents indicating the acute dryness of litter and biomass of the forested strand (Sati and Juyal, 2016). The LST reflects the complex interaction between the different factors like temperature, precipitation, socio-economic status, etc. (Sannigrahi et al., 2017a, b). The availability of resources and good quality of life in comparison to other districts led to the migration of the people from different areas to Dehradun Uttarkashi, Pauri Garhwal, and Nainital districts (Nandy et al., 2011). The migration of people led to the conversion of the forest area to an agricultural or residential area that in turn added to the net increase in the LST. The estimated LST of the individual fire days witnessed a sharp temperature increase in comparison to the non-fire days which reveals a high fire risk (in Uttarakhand) could be due to the subtle water stress condition (Chuvieco and Congalton, 1989; Guangmeng and Mei, 2004).

Different indicators (NDVI, EVI, and SAVI) document the drastic change in the vegetation phenology, depletion is significant for Uttarkashi, Pauri Garhwal, Dehradun, Almora and Tehri-Garhwal districts, whereas, Rudraprayag, Udham Singh Nagar, Bageshwar, Chamoli,

Champawat, Haridwar, and Pithoragarh districts reveal a positive change (**Table. S2**; **see supplementary file**). The observed phenological changes in the study area might have resulted because of the extreme surface moisture and temperature limiting condition that prevail in this region which might have triggered the functional changes in leaf foliage and wide-scale tree mortality (Chuvieco et al., 2004; Bartsch et al., 2009). The phenological disturbance is mostly associated with light use efficiency and absorbed/fractional photosynthetic capacity of a plant (Xiao et al., 2004). It favours the temperature and moisture limiting condition in an ecosystem that is detrimental to ideal photosynthesis and plant respiration (Yuan et al., 2015). Additionally, the seasonality and intensity of forest fire (crown, surface, and ground fires) have significantly controlled the phenological state and crown fuel structure, load, and moisture content of a forest by determining the seed or vegetation reproductive capacity and hence dismantle the native ecosystem structure and function meticulously (Flannigan et al., 2000). The season factor therefore can connect to the availability deciduous cover, which would regulate the warming and cooling behaviour of surface and ground fuel from direct sunlight especially during the summer time (Hely et al., 2000a, b). This could be a possible reason for the regular and periodical fire events happening over the extensive portions of the Himalayan region, specifically in the state of Uttarakhand and Himachal Pradesh, along with several other factors and drivers (which are not explicitly discussed in this research). Nevertheless, pursues special consideration from a researcher, ecologist, environmentalist, botanist and, a biologist to vividly explore/investigate the fire behaviour of this region to maintain the rich ecological and natural diversity of the Himalayan ecosystem.

Considering the vegetation indices, i.e., EVI, (**Fig. S1a, b)** and SAVI (**Fig. S1i, j)**, forest fires are mostly concentrated in the zone with high spatial changes of EVI and SAVI. However, SAVI is more capable of explicitly defining the spatiality of forest fire location, with maximum differences between the pre-fire and fire years. This is due to the addition of soil-adjustment factor (L) in its formulae, which reduces the background attenuation from soil and enhances the vegetation signal in the spectrum (Huete, 1988; Harris et al., 2011). The analysis revealed that LST is spatially related to the forest fire locations (**Fig. S1 c, d**). Among the selected burn indices, the spatial correlation between the actual occurrence of forest fires and changes of severity indices are found maximum using NBR (**Fig. S1 e, f**), followed by ΔLST/ΔEVI (**Fig. S1 i**), NMDI (**Fig. S1 g, h**) and GEMI (**Fig. S1 k**), respectively. The burn indices distinctly vary over the burned pixels and the changes between the pre-fire and fire years were found statistically significant **(Table. 2)**. Among all the experimented indices, EVI, SAVI, ΔLST/ΔEVI, exhibited a more spatially consistent association with NPP, than GEMI, NMDI, and LST, respectively (**Fig. S1: See supplementary**). The ΔLST and ΔEVI have demarcated the burned areas and offered a high spatial coherency between the satellite-derived burn delineation and actual forest fire locations. A similar observation has been made in the Iberian Peninsula, where maximizing the surface brightness temperature was found the most critical criterion for burn area delineation and mapping, instead of NDVI and other biophysical controls which cannot be directly used for burn scar delineation (Chuvieco et al., 2005).

**Table. S2** Variation of the selected burn severity indices in the pre-fire and forest fire years of different districts in the study area.

| Districts | Indices | Prefire | | | Fire | | |
|---|---|---|---|---|---|---|---|
| | | Mean | Min | Std.Dev | Mean | Min | Std.Dev |
| Rudraprayag | NBR | 0.4 | -0.12 | 0.25 | 0.36 | -0.08 | 0.19 |
| | BAI | 20.75 | 1.44 | 11.36 | 17.04 | 1.33 | 11.65 |
| | GEMI | 0.28 | -0.89 | 0.7 | 0.31 | -0.69 | 0.52 |
| | EVI | 0.21 | -0.2 | 0.17 | 0.47 | -0.09 | 0.27 |
| | NDVI | 0.4 | -0.14 | 0.29 | 0.47 | -0.09 | 0.27 |
| | SAVI | 0.22 | -0.09 | 0.16 | 0.21 | -0.11 | 0.16 |
| | NMDI | 0.58 | 0.31 | 0.17 | 0.56 | 0.35 | 0.14 |
| Tehri Garhwal | NBR | 0.3 | -0.27 | 0.2 | 0.25 | -0.17 | 0.19 |
| | BAI | 10.22 | 85.33 | 1.29 | 28.79 | 1.42 | 15.72 |
| | GEMI | 0.41 | -0.51 | 0.94 | 0.48 | -0.35 | 0.28 |
| | EVI | 0.25 | -0.2 | 0.11 | 0.25 | -0.2 | 0.11 |
| | NDVI | 0.49 | -0.14 | 0.18 | 0.47 | -0.2 | 0.18 |
| | SAVI | 0.26 | -0.09 | 0.1 | 0.25 | -0.06 | 0.11 |
| | NMDI | 0.49 | -0.37 | 0.1 | 0.51 | 0.01 | 0.1 |
| Udham Singh Nagar | NBR | 0.17 | 0.13 | 0.11 | 0.15 | -0.16 | 0.15 |
| | BAI | 6.16 | 59.99 | 9.64 | 21.8 | 9.52 | 6.74 |
| | GEMI | 0.51 | 0.23 | 0.05 | 0.53 | 0.25 | 0.06 |
| | EVI | 0.21 | -0.08 | 0.06 | 0.24 | -0.06 | 0.08 |
| | NDVI | 0.34 | -0.2 | 0.1 | 0.38 | -0.15 | 0.11 |
| | SAVI | 0.22 | -0.04 | 0.05 | 0.24 | -0.02 | 0.06 |
| | NMDI | 0.51 | 0.26 | 0.06 | 0.58 | 0.2 | 0.1 |
| Almora | NBR | 0.2 | -0.11 | 0.12 | 0.16 | -0.17 | 0.11 |
| | BAI | 7.1 | 73.33 | 11.14 | 29.54 | 9.53 | 10.44 |
| | GEMI | 0.52 | 0.4 | 0.03 | 0.52 | 0.33 | 0.05 |
| | EVI | 0.24 | 0.09 | 0.05 | 0.24 | 0.09 | 0.06 |
| | NDVI | 0.45 | 0.16 | 0.09 | 0.42 | 0.16 | 0.08 |
| | SAVI | 0.25 | 0.13 | 0.04 | 0.24 | 0.07 | 0.05 |
| | NMDI | 0.48 | 0.35 | 0.05 | 0.49 | 0.35 | 0.06 |
| Bageshwar | NBR | 0.33 | -0.09 | 0.22 | 0.29 | -0.21 | 0.19 |
| | BAI | 11 | 71.36 | 1.27 | 22.04 | 1.36 | 11.47 |
| | GEMI | 0.24 | -0.32 | 0.7 | -0.02 | -0.64 | 0.84 |
| | EVI | 0.21 | -0.2 | 0.12 | 0.25 | -0.2 | 0.12 |
| | NDVI | 0.43 | -0.13 | 0.22 | 0.45 | -0.09 | 0.21 |
| | SAVI | 0.22 | -0.1 | 0.11 | 0.22 | -0.12 | 0.13 |
| | NMDI | 0.53 | 0.29 | 0.15 | | | |
| Chamoli | NBR | 0.47 | -0.21 | 0.29 | 0.4 | -0.21 | 0.27 |
| | BAI | 18.15 | 1.13 | 14.49 | 19.39 | 1.28 | 15.81 |
| | GEMI | 0.5 | 0.4 | 0.21 | 0.41 | 0.31 | 0.18 |
| | EVI | 0.13 | -0.2 | 0.17 | 0.16 | -0.2 | 0.18 |
| | NDVI | 0.27 | -0.14 | 0.29 | 0.31 | -0.11 | 0.29 |
| | SAVI | 0.13 | -0.19 | 0.16 | 0.15 | -0.2 | 0.16 |
| | NMDI | 0.65 | 0.24 | 0.2 | 0.63 | 0.05 | 0.19 |
| Champawat | NBR | 0.2 | -0.14 | 0.14 | 0.17 | -0.12 | 0.13 |
| | BAI | 10.22 | 95.26 | 11.2 | 27.66 | 2.7 | 8.26 |
| | GEMI | 0.51 | 0.36 | 0.05 | 0.54 | -0.63 | 0.06 |
| | EVI | 0.24 | 0.03 | 0.06 | 0.26 | 0.05 | 0.06 |
| | NDVI | 0.47 | 0.06 | 0.11 | 0.47 | 0.09 | 0.1 |
| | SAVI | 0.25 | 0.05 | 0.05 | 0.26 | 0.03 | 0.06 |
| | NMDI | 0.47 | 0.35 | 0.05 | 0.49 | 0.37 | 0.06 |
| Dehradun | NBR | 0.27 | -0.09 | 0.16 | 0.21 | -0.19 | 0.16 |
| | BAI | 26.9 | 8.04 | 9.45 | 26.28 | 2.68 | 10.23 |
| | GEMI | 0.52 | 0.28 | 0.07 | 0.54 | -0.63 | 0.06 |
| | EVI | 0.26 | -0.01 | 0.08 | 0.25 | -0.07 | 0.09 |
| | NDVI | 0.47 | -0.02 | 0.13 | 0.44 | -0.15 | 0.14 |
| | SAVI | 0.25 | 0.07 | 0.06 | 0.25 | 0.01 | 0.08 |
| | NMDI | 0.49 | 0.27 | 0.06 | 0.52 | 0.31 | 0.07 |
| Haridwar | NBR | 0.17 | -0.16 | 0.11 | 0.12 | -0.21 | 0.09 |
| | BAI | 28.07 | 10.53 | 10.68 | 24.64 | 9.78 | 9.72 |
| | GEMI | 0.51 | 0.34 | 0.06 | 0.53 | 0.31 | 0.05 |
| | EVI | 0.22 | -0.03 | 0.05 | 0.23 | -0.06 | 0.06 |
| | NDVI | 0.38 | -0.07 | 0.09 | 0.37 | -0.14 | 0.09 |
| | SAVI | 0.22 | 0.03 | 0.05 | 0.23 | 0 | 0.05 |
| | NMDI | 0.54 | 0.33 | 0.08 | 0.58 | 0.36 | 0.08 |
| Nainital | NBR | 0.23 | -0.21 | 0.16 | 0.19 | -0.2 | 0.14 |
| | BAI | 27.53 | 8.32 | 9.76 | 26.96 | 3.26 | 11.02 |
| | GEMI | 0.54 | 0.36 | 0.06 | 0.54 | -0.15 | 0.07 |
| | EVI | 0.25 | 0.02 | 0.07 | 0.25 | 0.06 | 0.07 |
| | NDVI | 0.47 | 0.05 | 0.12 | 0.47 | 0.09 | 0.11 |
| | SAVI | 0.26 | 0.06 | 0.06 | 0.27 | 0.01 | 0.07 |
| | NMDI | 0.48 | 0.28 | 0.06 | 0.5 | 0.3 | 0.07 |
| Pauri Garhwal | NBR | 0.22 | -2.52 | 0.14 | 0.15 | -0.13 | 0.13 |
| | BAI | 29.11 | 9.32 | 9.84 | 30.42 | 6.92 | 13.29 |
| | GEMI | 0.52 | 0.1 | 0.06 | 0.52 | 0.23 | 0.07 |
| | EVI | 0.24 | -0.1 | 0.06 | 0.23 | -0.09 | 0.08 |
| | NDVI | 0.46 | -0.2 | 0.11 | 0.42 | -0.19 | 0.11 |
| | SAVI | 0.25 | -0.07 | 0.06 | 0.25 | -0.05 | 0.07 |
| | NMDI | 0.47 | -0.35 | 0.06 | 0.5 | 0.31 | 0.06 |
| Pithoragarh | NBR | 0.43 | -0.35 | 0.31 | 0.36 | -0.26 | 0.28 |
| | BAI | 22.84 | 1.17 | 17.64 | 20.68 | 1.25 | 15.84 |
| | GEMI | 0.11 | 0.04 | 0.19 | 0.07 | -0.08 | 0.04 |
| | EVI | 0.12 | -0.2 | 0.15 | 0.17 | -0.2 | 0.15 |
| | NDVI | 0.26 | -0.18 | 0.28 | 0.33 | **-0.12** | 0.28 |
| | SAVI | 0.13 | -0.23 | 0.14 | 0.15 | 0.61 | -0.17 |
| | NMDI | 0.64 | 0.25 | 0.2 | 0.62 | 0.22 | 0.18 |
| Uttarkashi | NBR | 0.36 | -0.08 | 0.19 | 0.39 | -0.26 | 0.28 |
| | BAI | 17.62 | 0.99 | 14.11 | 19.61 | 1.23 | 15.37 |
| | GEMI | 0.31 | -0.69 | 0.52 | 0.13 | -0.44 | 0.81 |
| | EVI | 0.26 | -0.2 | 0.17 | 0.17 | -0.2 | 0.17 |
| | NDVI | 0.47 | -0.09 | 0.27 | 0.33 | -0.16 | 0.3 |
| | SAVI | 0.21 | -0.11 | 0.16 | 0.15 | -0.13 | 0.16 |
| | NMDI | 0.56 | 0.35 | 0.14 | 0.65 | 0.28 | 0.19 |

Among the selected vegetation and burn indices, EVI, NDVI, and NBR were found most suitable and spatially coherent burn indicators, as it primarily provides more rigorous feasibility of burnt area mapping coupled with the field-based observation. Several studies have advocated the use of delta NBR to produce spatially explicit burn area maps, often referred as Burn Area Reflectance Classification (BARC) for delineating post-fire scars as it found to be reasonably correlated with the field based burn scar assessment (Keeley et al., 2009; Roy et al., 2006).

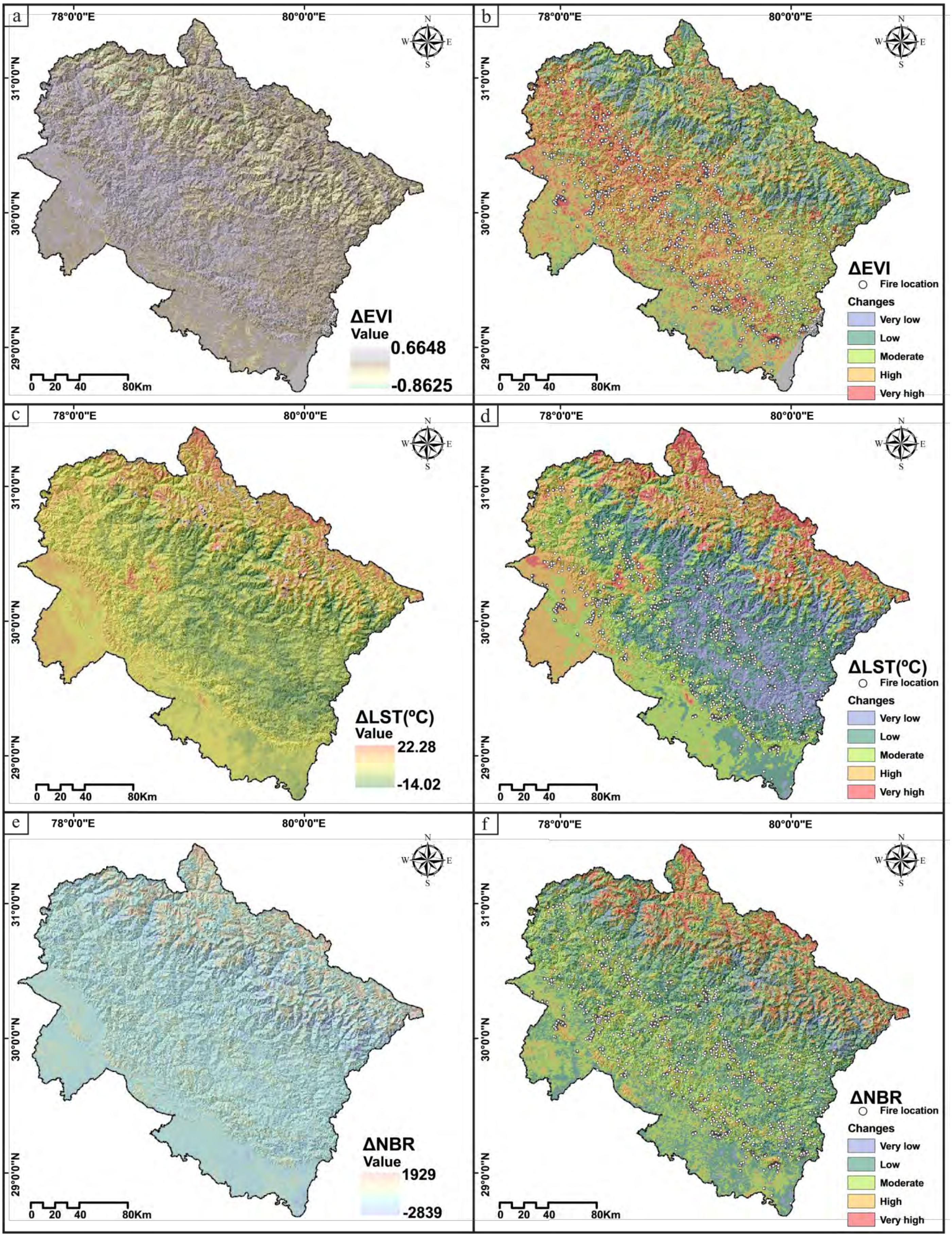

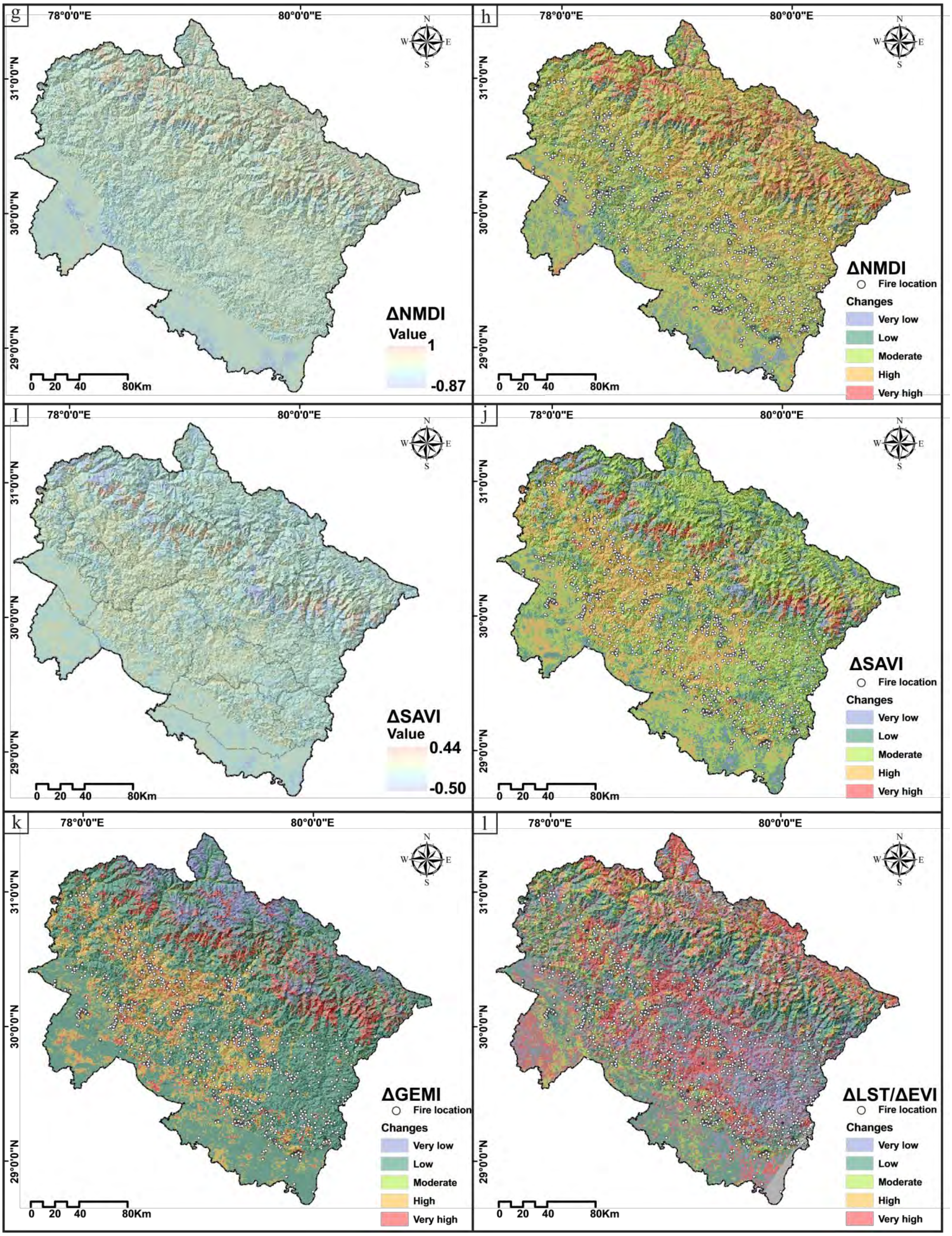

**Fig. S1:** The temporal variation and spatial segmentation of the selected burn severity indices viz. (a) & (b) ΔEVI, (c) & (d) ΔLST, (e) & (f) ΔNBR, (g) & (h) ΔNM I, (i) & (j) ΔSAVI, (k) ΔGEMI and (l) ΔLST/ΔEVI for the pre-fire and fire years, respectively.

**Table. 2** Student's 't' test showing mean differences of the burn severity indices and NPP between pre-fire and fire year.

| Model | Burn indices | Paired differences | | | t | df | Sig. (2-tailed) |
|---|---|---|---|---|---|---|---|
| | | Mean | Std. Deviation | Std. Error mean | | | |
| Pair 1 | EVI PREFIRE - EVI FIRE | 0.101 | 0.068 | 0.002 | 67.020 | 2047 | 0.000 |
| Pair 2 | GEMI PREFIRE - GEMI FIRE | 0.014 | 0.046 | 0.001 | 13.682 | 2047 | 0.000 |
| Pair 3 | LST PREFIRE - LST FIRE | -3.219 | 5.289 | 0.117 | -27.541 | 2047 | 0.000 |
| Pair 4 | NBR PREFIRE - NBR FIRE | 0.105 | 0.070 | 0.002 | 68.138 | 2047 | 0.000 |
| Pair 5 | NDVI PREFIRE - NDVI FIRE | 0.131 | 0.079 | 0.002 | 75.159 | 2047 | 0.000 |
| Pair 6 | NMDI PREFIRE - NMDI FIRE | -0.042 | 0.050 | 0.001 | -38.215 | 2047 | 0.000 |
| Pair 7 | SAVI PREFIRE - SAVI FIRE | 0.018 | 0.036 | 0.001 | 22.365 | 2047 | 0.000 |
| Pair 8 | NPP PREFIR - NPP FIRE | 14.521 | 13.064 | 0.289 | 50.302 | 2047 | 0.000 |

However, several noise factors, i.e. atmospheric contaminations, aerosols, bidirectional reflectance variation, clouds are often perturbs the remote sensed post-fire measured reflectance, makes the system insensitive to capture the post-fire changes, ultimately hinders the optimal uses of these burn indices for describing the physical shift in interest (Roy et al., 2006). The result exhibits good coherence between the spatial and temporal distribution of the selected burn indices and the intensity of forest fires. Therefore, the normal (positive) relation between the burn indices with the intensity of forest fires in a particular district is justified as the spatial agreement between the active fire locations and the delta burn-indices indicates the robust feasibility and practical applicability of satellite-based observation for active fire distribution (**Fig. S1: See supplementary**).

### *3.2 Impact of forest fire on ecosystem production and greenhouse gas discharge*

Forest fires are primary causative and natural drivers of biodiversity loss; depletion of terrestrial ecosystem productivity; forest carbon stocks; the decline of soil fertility and subsequent crop production; escalation of air pollutants and, increase in the magnitude of landslide susceptibility (e.g., Amiro et al., 2000; 2001; Verma & Jayakumar, 2012). The Uttarakhand state has hilly topography, and majority population (apart from plain) solely depend on the limited natural resources for fodder, medicinal plant, timber and others primary activities, which catalyzes the environmental and ecological degradation of this region (Nandy et al., 2011).

The NPP is widely used ecosystem indicator that assess the capacity of an ecosystem to act as a carbon source or sink and to track the unprecedented modifications in different biomes (Potter et al., 1993; Field et al., 1995). The biomass production and ecosystem services have a positive relationship with NPP, and the results corroborate the same, i.e., 146 gC $m^{-2}$ $month^{-1}$ and 142 gC $m^{-2}$ $month^{-1}$ in the non-fire year and fire year, derived from CASA NPP. The maximum change in NPP (gC $m^{-2}$ $month^{-1}$) have been observed in the districts having higher forest fire density and vice versa (**Fig. 3**). Among the 13 districts, the mean NPP decreased substantially in Champawat, followed by Nainital, Pauri-Garhwal, Dehradun, Tehri-Garhwal, Almora and Udham-Singh Nagar, respectively. Whereas, the remaining six districts exhibit an increase of mean NPP with the highest value in Rudraprayag, followed by Uttarkashi, Pithoragarh, Chamoli, Bageshwar and Haridwar districts, respectively (**Fig. 4 a**). The total NPP declined drastically in the districts having higher forest fire density viz. highest in Pauri-Garhwal (255 kg C $month^{-1}$), followed by Nainital (214 kg C $month^{-1}$), Champawat (136 kg C $month^{-1}$), Tehri-Garhwal (104 kg C $month^{-1}$), Dehradun (100 kg C $month^{-1}$), Udham Singh Nagar (91 kg C $month^{-1}$) and Almora (57 kg C $month^{-1}$), respectively (see **Fig. 4 a**). District-wise ecosystem production (NPP, GPP, and Biomass) and efficiency (ELUE) of the pre and fire years are shown in **Fig. 4b**. Maximum values of NPP, GPP, Biomass, and ELUE were observed in the pre and fire years varied highly in Champawat, followed by Nainital, Tehri-Garhwal, Dehradun, Pauri-Garhwal, Almora and Udham Singh Nagar districts respectively and the remaining districts reflect minimum to negligible changes (**Fig. 4b**).

To understand the effect of wildfires on the carbon budget the accurate measurement of fire intensity and level of fire severity of each plant types and biomes is essential because the volatilization and redistribution of carbon due to active forest fires depends on type and level of fire severity (Wang et al., 2001). **Fig. 5** exhibits a strong coherence between FFD and ΔNPP ($R^2$ = 0.75, RMSE = 5.03gC $m^{-2}$ $month^{-1}$), highest in Nainital, followed by Champawat, Almora, Pauri-Garhwal, Dehradun, Tehri-Garhwal and Udham Singh Nagar, respectively. The districts having lower FFD exhibits minimum changes of the NPP viz. Rudraprayag, Uttarkashi, Pithoragarh, and Haridwar, respectively (**Table. S3; see supplementary file**). The Δ NPP is very high in Champawat, Nainital, and Pauri-Garhwal, high in Champawat, Nainital, and Bageshwar, moderate in Udham Singh Nagar and Haridwar, low in Chamoli and Bageshwar and very low in Uttarkashi, Rudraprayag, and Pithoragarh (**Fig. 6).** This can be attributed to the high forest fire density and associated forest cover losses due to 2016 forest fire. A similar observation has been made for the Boreal forest region (Canada), where measured $CO_2$ flux from eddy covariance and LUE modelled NPP shows that the forest fire has reduced the net downward fluxes of carbon, however, it (carbon flux) has been increased 10-30 years after the fire event (Peng et al., 1999; Amiro et al., 2000; Amiro et al., 2003). Amiro et al., (1999) study on the measurements of net carbon flux over the boreal forest have revealed that fire disturbance disrupted the overall carbon cycle at any given ecosystem, and it needs 15 to 30 years following a fire event to reach the normal photosynthetic level, which appears to be a significant entity to any carbon balance model. However, several additional attributes including decomposition process, heterotrophic respiration, etc. are required for efficient carbon budget and flux estimation. These approximations have not covered in this research could be a future scope of this work. In the Himalayan region, strong negative impact of forest fire on ecosystem productivity, soil nutrient status (soil organic carbon, nitrogen, phosphorus and potassium), and understorey vegetation structure can be controlled by educating local villagers about the adverse effects of active forest fire (both human-induced and natural) on their native ecosystem (Kumar et al., 2013).

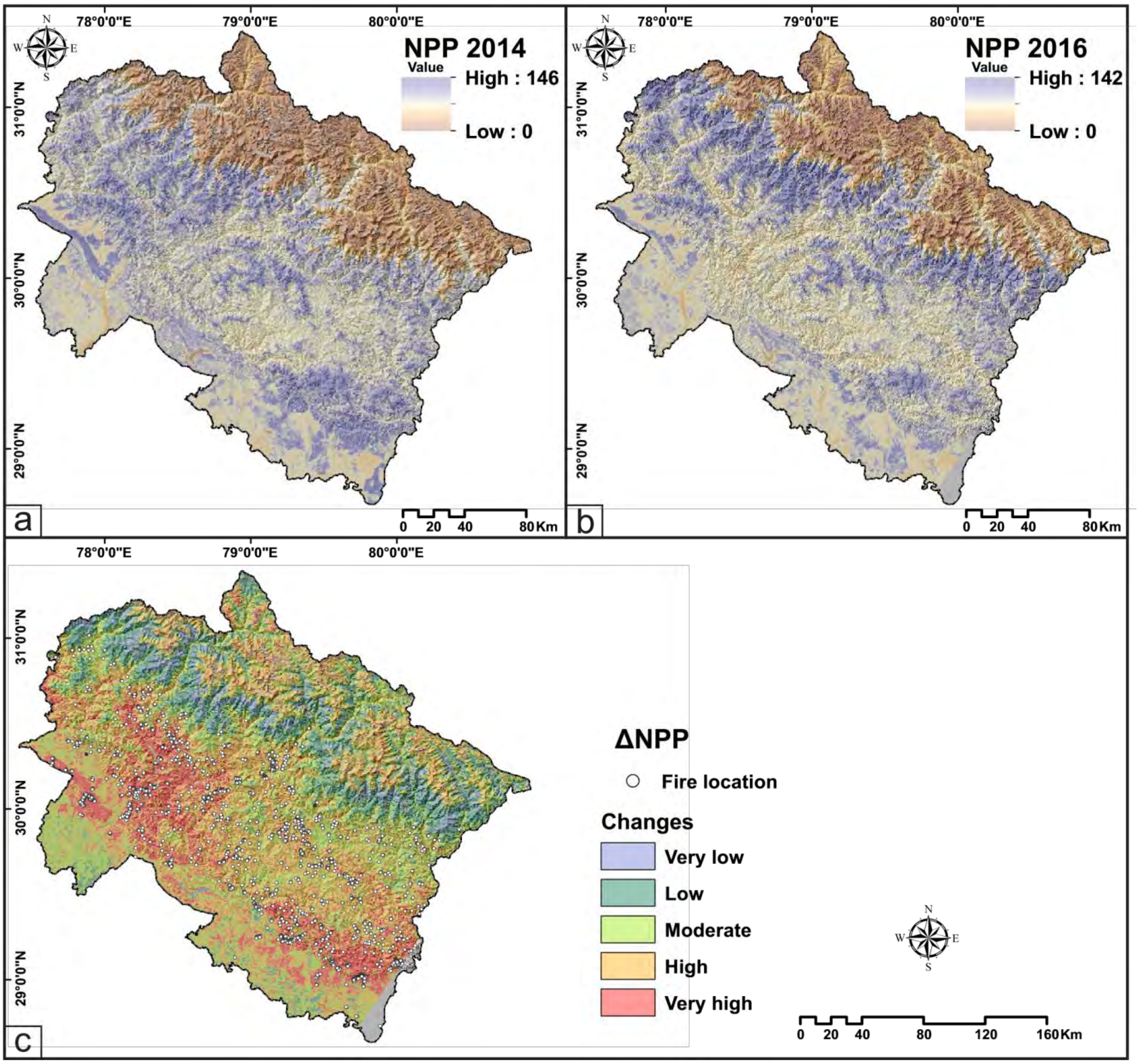

**Fig. 3:** Distribution of estimated NPP for (a) pre-fire (2014) and (b) fire year (2016), and (c) the resultant change of NPP between pre-fire (2014) and fire year (2016) in the study region with forest fire locations.

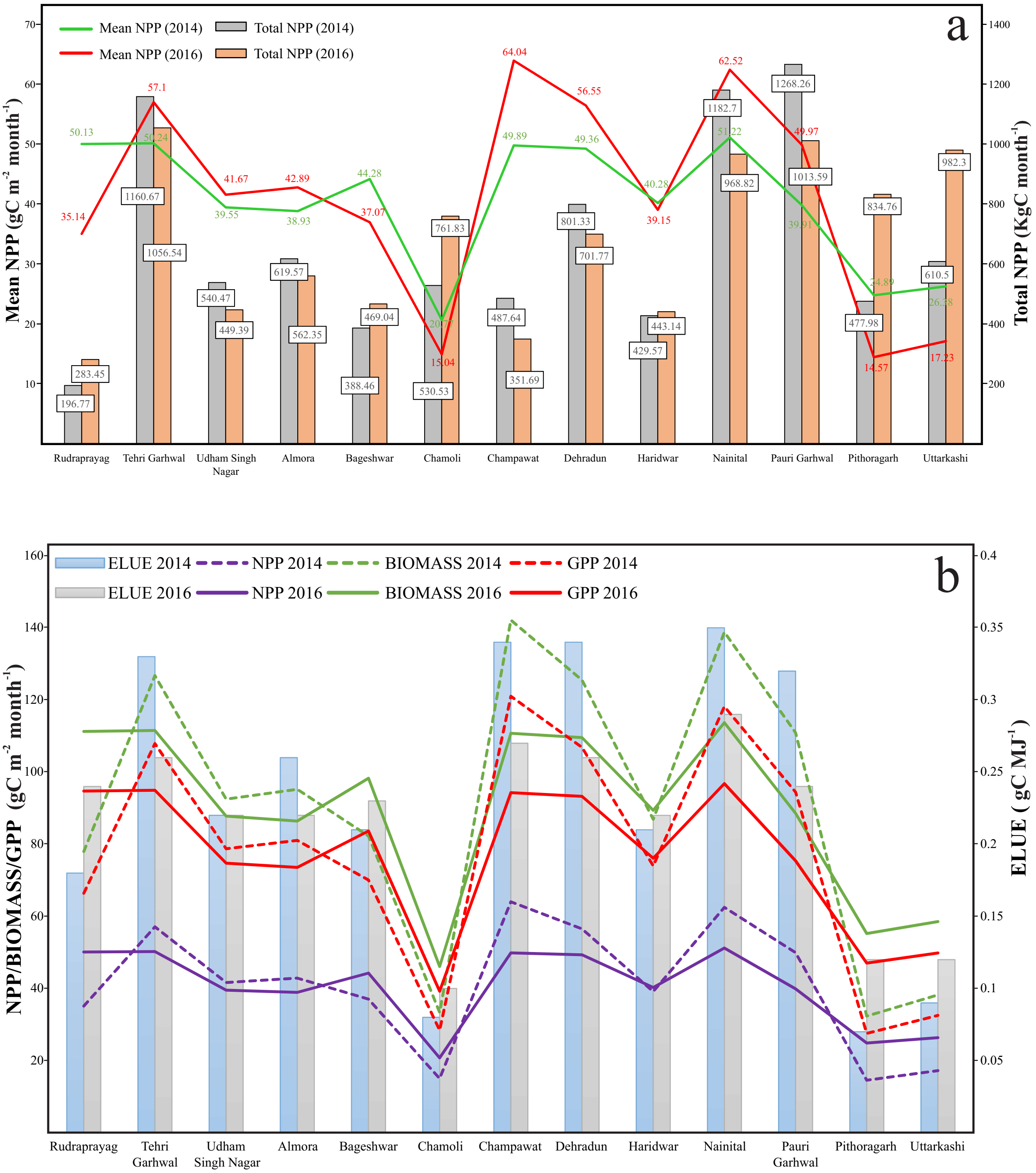

**Fig. 4:** (a) Statistical summary of change in mean and total NPP (gC $m^{-2}$ $month^{-1}$) between the pre-fire and forest fire years in the study region, (b) the status of NPP (gC $m^{-2}$ $month^{-1}$), Biomass (gC $m^{-2}$ $month^{-1}$), GPP (gC $m^{-2}$ $month^{-1}$), and ELUE (gC $MJ^{-1}$) in the pre-fire and forest fire years for different districts in the study area.

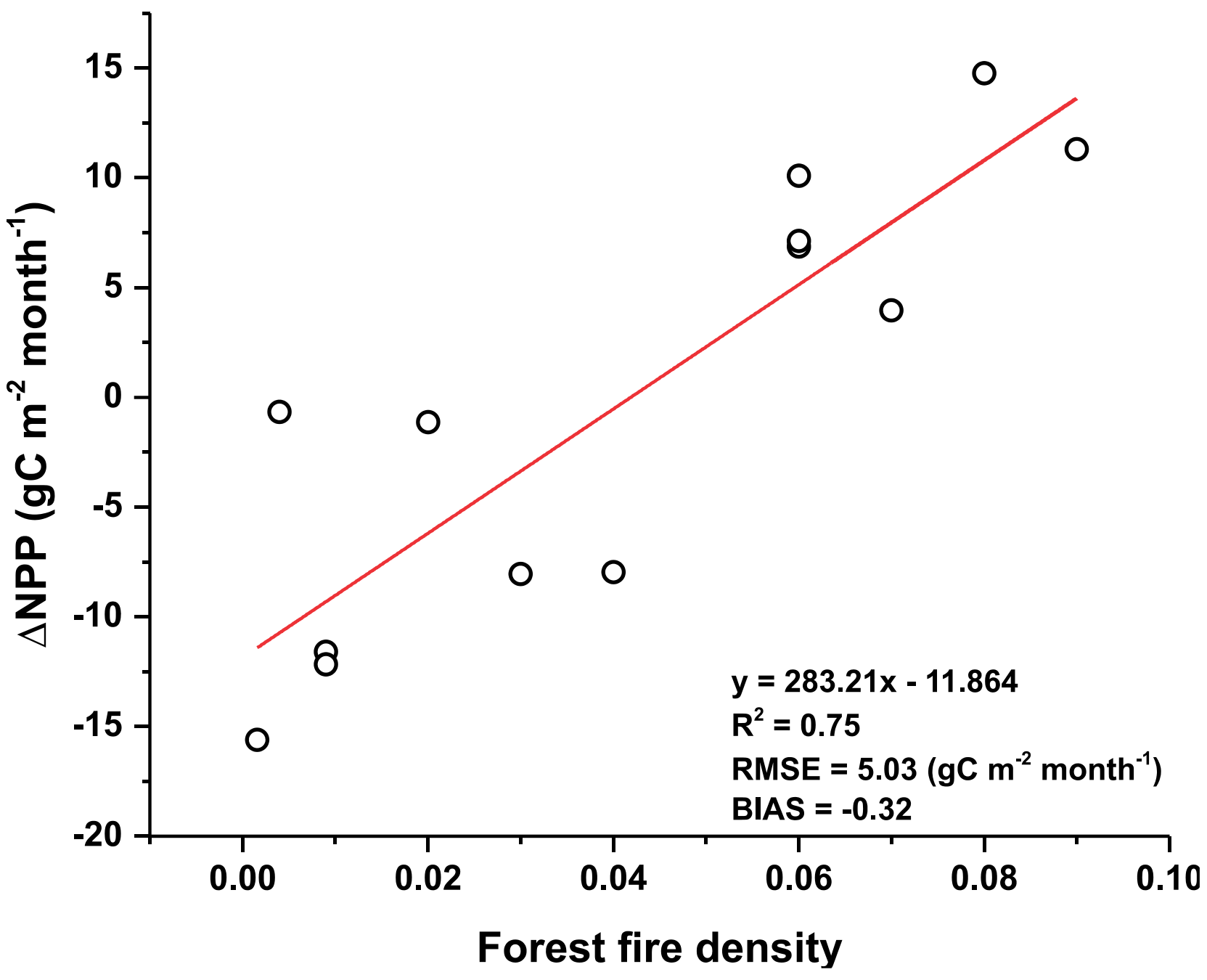

**Fig. 5:** Coefficient of determination between the Forest Fire e nsity and ΔNPP in the study region.

The relationship between the Δ burn indices (ΔEVI, ΔGEMI, ΔLST, ΔNBR, ΔNDVI, ΔSAVI and ΔNMDI) and ΔNPP is evaluated as the ecosystem productivity can be linked to these burn indices (**Table. 3)**. **Fig. 7** depicts spatial distribution and sensitivity between the Δburn

**Table. S3** District wise forest fire density and ΔNPP in the study area.

| Districts | Forest Fire Density | Δ NPP (gC $m^{-2}$ $month^{-1}$) |
|---|---|---|
| Rudraprayag | 0.0016 | -15.60 |
| Tehri Garhwal | 0.06 | 6.86 |
| Udham Singh Nagar | 0.004 | 0.66 |
| Almora | 0.07 | 3.96 |
| Bageshwar | 0.04 | -7.96 |
| Chamoli | 0.03 | -8.05 |
| Champawat | 0.08 | 14.76 |
| Dehradun | 0.06 | 7.11 |
| Haridwar | 0.02 | -1.13 |
| Nainital | 0.09 | 11.30 |
| Pauri Garhwal | 0.06 | 10.08 |
| Pithoragarh | 0.009 | -11.59 |
| Uttarkashi | 0.009 | -12.17 |

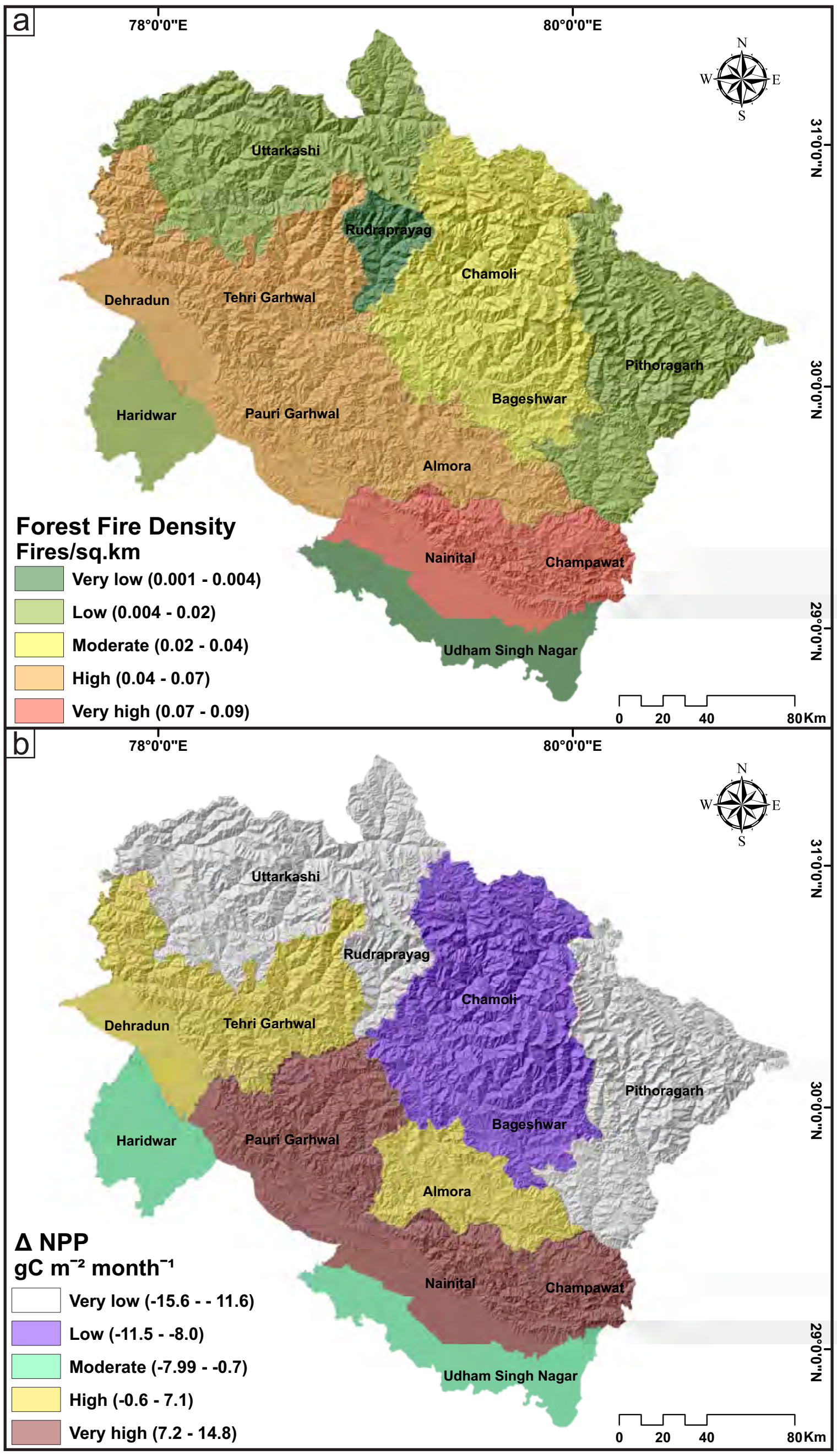

**Fig. 6:** ist rict wise spatial coherence of (a) Forest Fire ensity (FF ) and, (b) ΔNPP in the study region.

**Table. 3** Parametric (Pearson correlation coefficient) and Spearman rank correlation analysis between delta burn indices and NPP.

| **Model** | **Burn indices** | **Pearson's r** | **Sig. (2-tailed)** | **N** | **Spearman's r** | **Sig. (2-tailed)** | **N** |
|---|---|---|---|---|---|---|---|
| Pair 1 | Δ EVI & Δ NPP | 0.536** | 0.000 | 2048 | 0.379** | 0.000 | 2048 |
| Pair 2 | Δ GEMI & Δ NPP | 0.156** | 0.000 | 2048 | 0.097** | 0.000 | 2048 |
| Pair 3 | Δ LST & Δ NPP | -0.14** | 0.000 | 2048 | -0.18** | 0.000 | 2048 |
| Pair 4 | Δ NBR & Δ NPP | 0.36** | 0.000 | 2048 | 0.23** | 0.000 | 2048 |
| Pair 5 | Δ NDVI & Δ NPP | 0.548** | 0.000 | 2048 | 0.406** | 0.000 | 2048 |
| Pair 6 | Δ NMDI & Δ NPP | -0.39** | 0.000 | 2048 | -0.31** | 0.000 | 2048 |
| Pair 7 | Δ SAVI & Δ NPP | 0.164** | 0.000 | 2048 | 0.101** | 0.000 | 2048 |

** Correlation is significant at the 0.01 level (2-tailed).

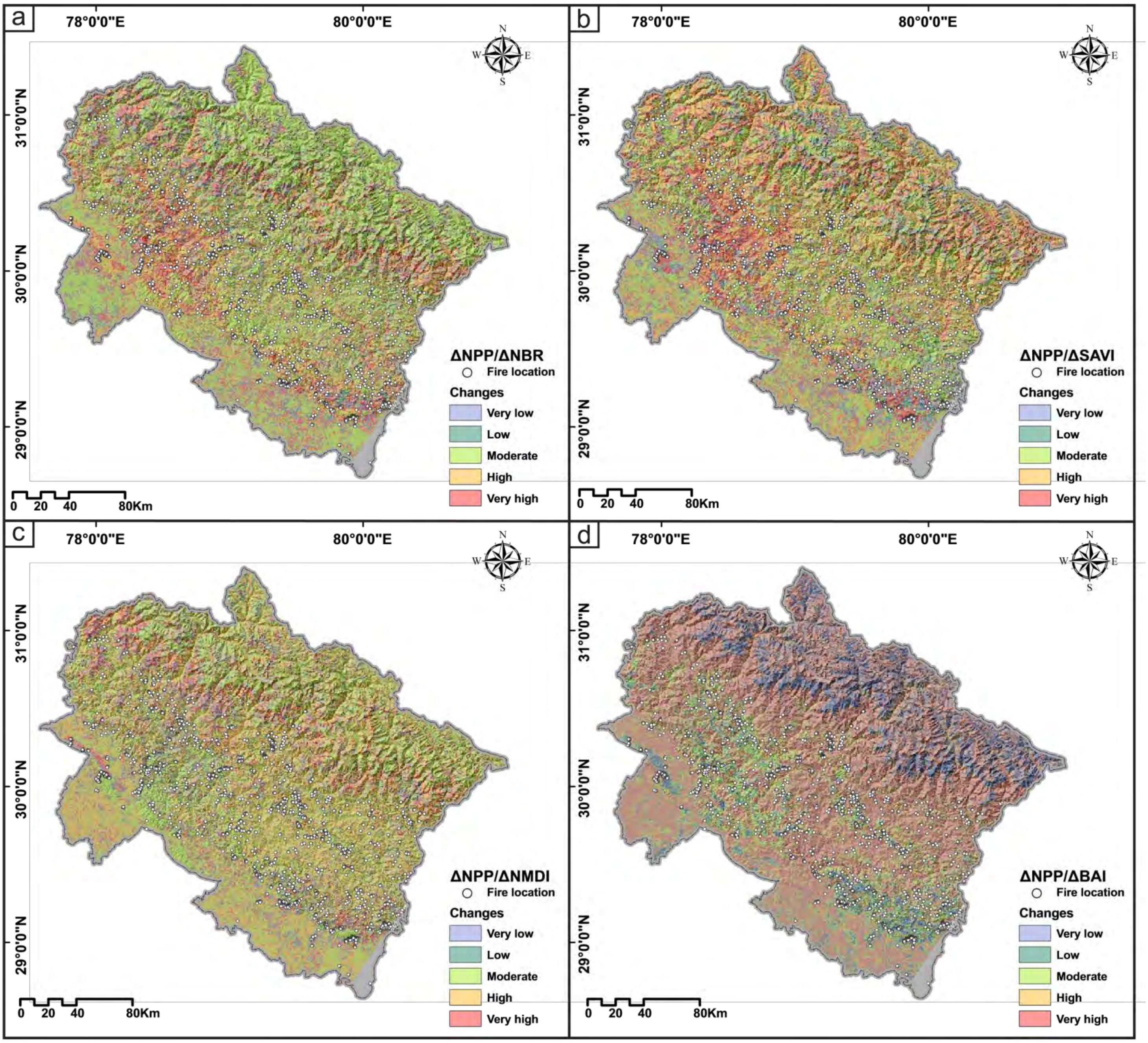

**Fig. 7:** **S**ensitivity between the (a) ΔNPP and ΔNBR, (b) ΔNPP and ΔSAVI, (c) ΔNPP and ΔNM I and (d) ΔNPP and ΔBAI in the study region.

indices and ΔNPP with active forest fire locations. Among all the burn indices, high spatial coherence has been observed between the ΔSAVI and ΔNPP (**Fig. 7b**), followed by ΔNBR and ΔNPP (**Fig. 7a**), ΔNMDI and ΔNPP (**Fig. 7c**), and ΔBAI and ΔNPP (**Fig. 7d**), respectively. All the correlation coefficients values between the Δ burn indices and Δ NPP are significant at 0.01 level (2-tailed). The values of the correlation coefficient between Δ NPP and different indices are highest for Δ NDVI (0.41) and lowest for Δ GEMI (0.1). Correlation coefficients for Δ LST and Δ NMDI are found a negative (**Table. 3**).

The carbon emissions and sequestration were quantified for 13 districts using estimated NPP (**Fig. 8)**. The mean NPP of the high FFD districts showed a drastic change in the forest fire year (2016) and, can be attributed to significant biomass loss and the resultant forest carbon stock due to forest fire (Fearnside, 2000; Gillett et al., 2004). The subsequent emissions of other greenhouse compounds from forest fire of the various districts in the study area are shown in **Table. 4**. We observed that the Nainital, Champawat and Tehri-Garhwal districts show high emission of $CO_2$ whereas Haridwar and Udham Singh Nagar show low carbon emissions (**Fig. 8)**. **Table. 4** shows the detail description of ΔBiomass (gC), emission of Carbon compounds (C, $CO_2$, $CH_4$, CO), Nitrogen compounds ($NO_2$, $NO_x$) and particulate matters of the 13 districts in Uttarakhand state. Maximum change in ΔBiomass has been observed in Champawat (31.41), followed by Nainital (25.09), Pauri-Garhwal (22.33), Dehradun (15.96), Tehri-Garhwal (15.23), Almora (8.79) and Udham Singh Nagar (4.71), respectively. However, the maximum negative change in ΔBiomass are observed in Rudraprayag (-33.28), Pithoragarh (-22.91), Uttarkashi (-20.31), Bageshwar (-16.01), Chamoli (-12.72) and Haridwar (-2.51) districts, respectively. These changes (total biomass) could be linked to post-fire mortality and associated changes as shown by de Vasconcelos et al., (2013) for South Western Brazilian Amazonia, which revealed a significant loss of total and above ground biomass due to the increase of large-scale tree mortality after 1st ($1.6 \times 10^6$ Mg and $1.4 \times 10^6$ Mg) and 4th year ($4.4 \times 10^6$ Mg and $3.7 \times 10^6$ Mg) following the fire event. The resultant emission of total and above ground carbon stock after the fire increased in subsequent year ($0.8 \times 10^6$ Mg C and $0.7 \times 10^6$ Mg C after 1st year and $2.2 \times 10^6$ Mg C and $1.8 \times 10^6$ Mg C after 4th year) depending on the balance between the rate of decomposition of dead tree and the regeneration of fresh canopy in a given period of time (e.g. de Vasconcelos et al., 2013). Therefore, post-fire mortality assessment is highly recommended to get the real insights about collective response/regeneration time of a plant community due to anomalous forest disturbance. The examined carbon and nitrogen compounds follow the similar trend as the Δbiomass in the mentioned districts (**Table. 4**). The overall response can be ascribed to the acute biomass burning (Andreae, 2001), land-use and land cover changes in different eco-regions for biofuels (Searchinger et al., 2008), severe deforestation and associated forest degradation (Van der Werf, et al., 2009). Among the 13 districts, the maximum net emissions of carbon and nitrogen compounds have been observed in 7 districts (accounting for high biomass and forest cover loss due to 2016 forest fire), whereas, the rest of the 6 districts act as the sequester of greenhouse compounds (**Table. 4**). The results echo the fact that reducing the fossil fuel emissions to the atmosphere and fire control activities are one of the essential elements for stabilizing atmospheric $CO_2$ concentration (Dixon & Krankina, 1993; Werf et al., 2009).

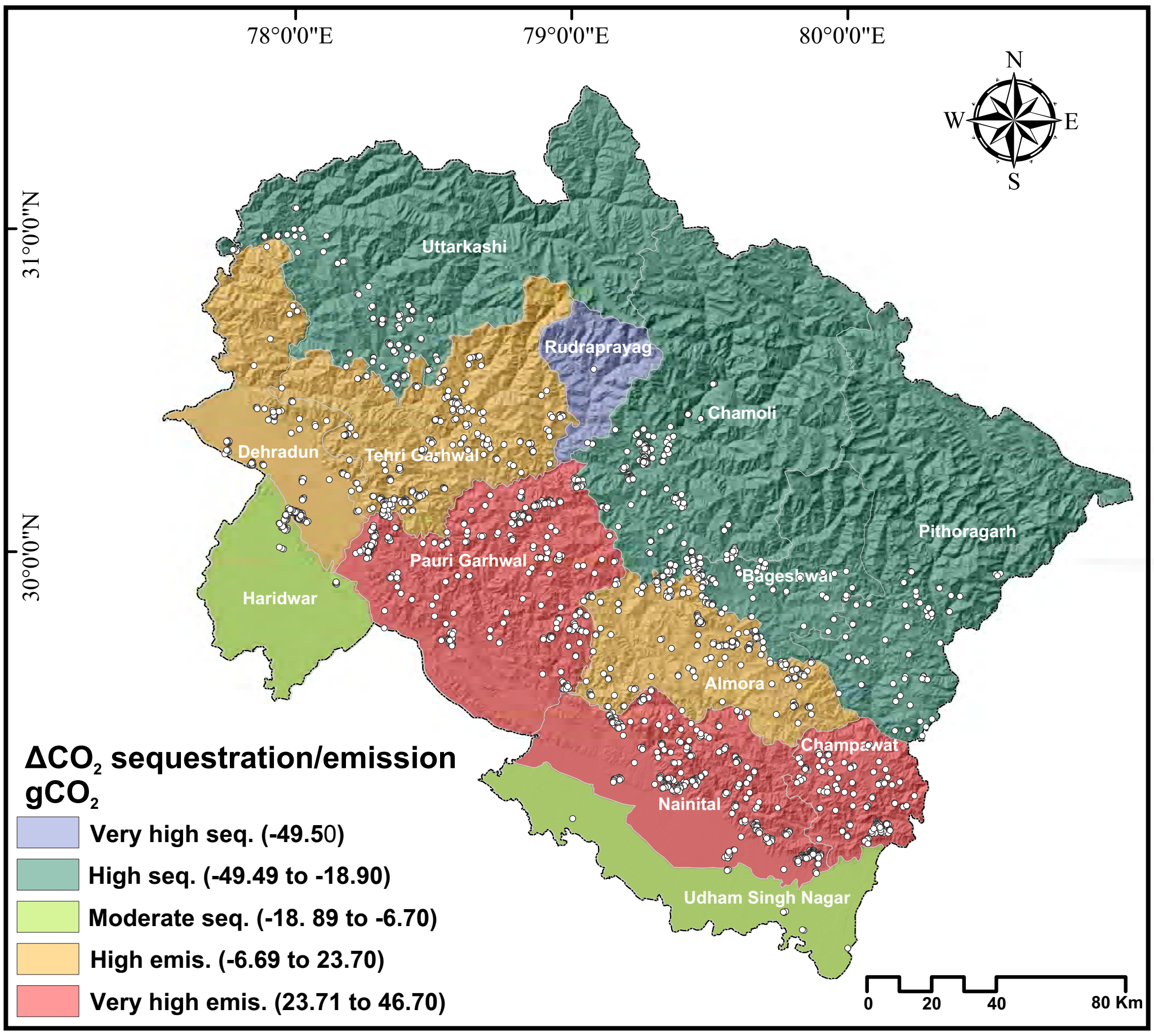

**Fig. 8:** Spatial distribution of the difference between the $CO_2$ emission and sequestration in the study area for pre-fire (2014) and fire year (2016).

**Table. 4** District wise summary of emission of the different greenhouse compounds in the study region.

| Districts | Δ Biomass | C | $CO_2$ | $CH_4$ | CO | $NO_2$ | $NO_x$ | Particulate matter |
|---|---|---|---|---|---|---|---|---|
| Rudraprayag | -33.28 | -13.48 | -49.46 | -0.22 | -3.14 | -0.0015 | -0.0035 | -0.10 |
| Tehri Garhwal | 15.23 | 6.17 | 22.64 | 0.10 | 1.44 | 0.00068 | 0.00158 | 0.05 |
| Udham Singh Nagar | 4.71 | 1.91 | 7 | 0.03 | 0.44 | 0.00021 | 0.00049 | 0.01 |
| Almora | 8.79 | 3.56 | 13.07 | 0.06 | 0.83 | 0.00039 | 0.00091 | 0.03 |
| Bageshwar | -16.01 | -6.48 | -23.79 | -0.10 | -1.51 | -0.0007 | -0.0017 | -0.05 |
| Chamoli | -12.72 | -5.15 | -18.91 | -0.08 | -1.20 | -0.0006 | -0.0013 | -0.04 |
| Champawat | 31.41 | 12.72 | 46.69 | 0.20 | 2.96 | 0.0014 | 0.00327 | 0.09 |
| Dehradun | 15.96 | 6.46 | 23.73 | 0.10 | 1.51 | 0.00071 | 0.00166 | 0.05 |
| Haridwar | -2.51 | -1.02 | -3.73 | -0.02 | -0.24 | -0.0001 | -0.0003 | -0.01 |
| Nainital | 25.09 | 10.16 | 37.29 | 0.16 | 2.37 | 0.00112 | 0.00261 | 0.07 |
| Pauri Garhwal | 22.33 | 9.04 | 33.19 | 0.14 | 2.11 | 0.00099 | 0.00232 | 0.07 |
| Pithoragarh | -22.91 | -9.28 | -34.05 | -0.15 | -2.16 | -0.001 | -0.0024 | -0.07 |
| Uttarkashi | -20.31 | -8.23 | -30.19 | -0.13 | -1.92 | -0.0009 | -0.0021 | -0.06 |

The Himalayan forest ecosystem has a significant role in the global terrestrial carbon balance to assimilate $CO_2$ from the atmosphere, storage of carbon and release of greenhouse components to the atmosphere (Kauppi et al., 1992; Dixon et al., 1993).The accelerated natural and human interventions have led frequent forest fires in Uttarakhand Himalayan region significantly dismantling the native ecosystem functions and posing a serious threat to its highly diverse and rich ecosystem. Conserving the biodiversity and natural resources of this ecosystem from any disruptive interventions and calamities should be the core of the policy making for a sustainable development of the region. Most of the mountainous societies thrive in close socio-ecological association with nature and have traditions, values, and faith weaved as the fundamental fabric of the indigenous cultures. Theses association often appear to be fruitful to maintain a fair forest management policies integrated with the active participation of local peoples through the Joint Forest Management (JFM) and community-based forest management (Anthwal et al., 2010; Dutta et al., 2012; Bhattacharya et al., 2010). Moreover, Sarin (2001) study had revealed that the active involvement and empowering women for managing and preserving of local forest resources through integrated public participatory and citizen science-based approaches having an impeccable impact on formulating and regulating effective forest management policies.

**Conclusion**

In this study, the impact of forest fire severity and its associated vegetation loss is evaluated by using remote sensing products and climate data in a data scarce, highly enriched mountainous ecosystem of Uttarakhand, India. The impact of forest fire on the NPP of the ecosystem was assessed by using several burn indices (NBR, BAI, SAVI, EVI, NDVI, GEMI, and NMDI). In summary, a good correlation coefficient (r) values between the ΔNPP vs. EVI (0.54); NDVI (0.55); NBR (0.36); NMDI ($r = -0.39$); SAVI (0.16), GEMI (0.16), and LST (-0.14) were observed. These correlation values collectively indicate the coherence of forest burn with the loss in terrestrial NPP. Among the burn indices (EVI, NDVI, GEMI, NBR, BAI, SAVI, and LST), EVI, NDVI, NBR, and NMDI were found to be most suitable and spatially coherent burn indicator. The ΔNPP and estimated FFD are correlated ($R^2 = 0.75$) and ranges between very high to moderate range in most of the districts.

Maximum changes in Biomass, $CO_2$, CO, $CH_4$, $NO_2$, $NO_X$, particulate matter have been observed in all the 13 districts of Uttarakhand, and they exhibit a similar trend and show a positive correlation with ΔNPP. Apart from this, the emissions of the greenhouse compounds were observed only for a particular month of a year, which might have caused an estimated bias for not accounting the post-fire emissions. This will be the idea for the future research. The newly introduced approach (ΔNPP /Δburn indices) exhibits a high potential in quantifying the loss in ecosystem productivity due to forest fires in different eco-regions around the world. In addition to this, the current approach also helps in an accurate delineation of burn areas using the remotely sensed data, which can be used in broader aspects if the more accurate field-based observation can be obtained in the near future. Therefore, the developed approach and the similar approach in this direction will be helpful for planning agencies, consultancies and local government in planning and management of different fire mitigation strategies in an area. The strong negative impact of forest fire on terrestrial ecosystem production in Uttarakhand can be controlled by educating local villagers about the adverse effects of active forest fire (both human-induced and natural) on their native ecosystem. A detail investigation (both quantitative and qualitative) is, therefore, essential for developing a fire inventories for different plant functional types in the Himalayan region to cope up with the ecological destruction and biodiversity losses due to the forest fire.

**Acknowledgement**

Authors are indebted to Bhumika Uniyal for her sincere support and input to enhance the quality of the research. SS acknowledge University Grant Commission (UGC) for providing continuous research fellowship to carry out the research at Indian Institute of Technology (IIT), Kharagpur, India. SB would like to acknowledge INSPIRE Fellowship Programme (Award Number: IF131138) funded by Department of Science & Technology (DST, New Delhi) for doctoral research being carried out at the Indian Institute of Technology (IIT), Kharagpur (India). SR thanks the Ministry of Human Resource Development (MHRD, New Delhi) for providing continuous research fellowship for doctoral work being carried out at the Indian Institute of Technology (IIT), Kharagpur (India). VR acknowledges Ministry of Earth Science (MoES) for research assistantship.

**Appendices**

***Estimation of Δ Land Surface Temperature***

$$\Delta LST = LST_{prefire} - LST_{fire} \tag{13}$$

***Estimation of vegetation dynamics***

$$EVI = 2.5 \times \frac{\rho_{NIR} - \rho_{RED}}{1 + \rho_{NIR} + 6 \times \rho_{RED} - 7.5 \times \rho_{BLUE}} \tag{14}$$

$$\Delta EVI = EVI_{prefire} - EVI_{fire} \tag{15}$$

$$NDVI = \frac{(\rho_{NIR} - \rho_{RED})}{(\rho_{NIR} + \rho_{RED})} \tag{16}$$

Where $\rho_{NIR}$, $\rho_{RED}$ and $\rho_{BLUE}$ are the near infrared, red and blue spectral bands reflectance of MODIS satellite imagery.

***Estimation of burn indices***

$$SAVI = (1+0.5) \times \frac{\rho_{NIR} - \rho_{\mathrm{Re}d}}{\rho_{NIR} + \rho_{\mathrm{Re}d} + 0.5} \tag{17}$$

$$\Delta SAVI = SAVI_{prefire} - SAVI_{fire} \tag{18}$$

$$NBR = \frac{\rho_{NIR} - \rho_{SWIR}}{\rho_{NIR} + \rho_{SWIR}} \tag{19}$$

$$\Delta NBR = (NBR_{prefire} - NBR_{fire}) \times 1000 \tag{20}$$

$$BAIM = \frac{1}{(\rho_{NIR,C} - \rho_{NIR})^2 + (\rho_{SWIR,C} - \rho_{SWIR})^2} \tag{21}$$

$$\Delta BAIM = BAIM_{prefire} - BAIM_{fire} \tag{22}$$

$$NMDI = \frac{R_{860nm} - (R_{1640nm} - R_{2130nm})}{R_{860nm} + (R_{1640nm} - R_{2130nm})} \tag{23}$$

$$\Delta NMDI = NMDI_{prefire} - NMDI_{fire} \tag{24}$$

$$GEMI = \frac{\eta(1-0.25\eta) - (\rho_{\mathrm{Re}d} - 0.125)}{(1-\rho_{\mathrm{Re}d})} \tag{25}$$

$$\eta = \frac{2(\rho^2_{\mathrm{NIR}} - \rho^2_{\mathrm{Re}d}) + 1.5\rho_{\mathrm{NIR}} + 0.5\rho_{\mathrm{Re}d}}{(\rho_{\mathrm{NIR}} + \rho_{\mathrm{Re}d} + 0.5)} \tag{26}$$

$$\Delta GEMI = GEMI_{prefire} - GEMI_{fire} \tag{27}$$

Where $\rho_{SWIR}$, shortwave infrared reflectance, $\rho_{NIR,C}$, NIR reflectance of convergence point, 0.05 $\rho_{SWIR,C}$, SWIR reflectance of convergence point, 0.2.

***VPM derived NPP***

$$GPP = APAR \times \varepsilon_{\max} \times T_s \times W_s \times P_s \tag{28}$$

where $APAR$ (MJ m$^{-2}$ year$^{-1}$) is the absorbed active photosynthetic radiation by a green canopy, $\varepsilon_{\max}$ is the maximum light use efficiency (gC MJ$^{-1}$) in the absence of down regulators scalars (Xiao et al., 2004), $T_s$, $W_s$ and $P_s$ are the downward regulator scalars vary between 0 – 1, denotes low to high environmental stress condition for vegetation photosynthesis. APAR was retrieved as follows

$$APAR = PAR \times fPAR \tag{29}$$

$$PAR = 0.5 \times iSR \tag{30}$$

$$fPAR = \gamma \times EVI \tag{31}$$

Where $PAR$ is the incoming shortwave active photosynthetic radiation (MJ m$^{-2}$ month$^{-1}$), $iSR$ is the incoming solar radiation (MJ m$^{-2}$ month$^{-1}$, $fPAR$ is the fraction of PAR which was estimated directly from the linear function of EVI (Xiao et al., 2004, 2005), where the coefficient $\gamma$ is set to 1 for simple parameterization of model function. The down regulator scalars were calculated as follows:

$$T_s = \frac{(T_a - T_{\min})(T_a - T_{\max})}{(T_a - T_{\min})(T_a - T_{\max}) - (T_a - T_{opt})^2} \tag{32}$$

$$W_s = \frac{1 + LSWI}{1 + LSWI_{\max}} \tag{33}$$

$$LSWI = \frac{\rho_{NIR} - \rho_{SWIR}}{\rho_{NIR} + \rho_{SWIR}} \tag{34}$$

$$P_s = \frac{(1 + LSWI)}{2} \tag{35}$$

Where $T_a, T_{\max}, T_{\min}, T_{opt}$ is the average, maximum, minimum and optimum temperature (ºC) respectively that controls the photosynthetic activity. $T_s$ is set to be zero if air temperature falls below the minimum temperature (Xiao et al., 2004). $W_s$ is the water regulator scalar and was retrieved from satellite-based land surface water index (LSWI) and $LSWI_{\max}$, maximum LSWI at the growing season, $\rho_{NIR}$ and $\rho_{SWIR}$ is the near infrared and shortwave infrared bands of optical spectrum, $P_s$ is the effect of leaf age on a photosynthetic canopy.

***CASA Model derived NPP***

$$NPP = PAR \times fPAR \times \varepsilon_{\max} \times T_{S_1} \times T_{S_2} \times W_S \tag{36}$$

The biophysical variable $fPAR$ (Rahman et al., 2004) is calculated as follows:

$$fPAR = 1.24 * NDVI - 0.168 \tag{37}$$

Where NDVI is the normalized difference vegetation index, $T_{s_1}, T_{s_2}, W_s$, are the temperature related stress scalars and water stress scalar downward the potential LUE from optimum was retrieved as follows:

$$T_{s_1} = 0.8 + 0.02 \times T_{opt} - 0.0005 \times (T_{opt})^2 \quad (38)$$

$$T_{s_2} = \frac{1.1919}{\{1 + e^{[0.2(T_{opt} - 10 - T)}\} / \{1 + e^{[0.3(-T_{opt} - 10 + T)}\}} \quad (39)$$

Where $T$ is the average temperature for a particular month. Here we have adopted the VPM $W_s$ approach to estimate the water stress scalar which had illustrated earlier (Yuan et al., 2015).